\documentclass[
english,			% idioma adicional para hifenização
french,				% idioma adicional para hifenização
spanish,			% idioma adicional para hifenização
brazil,				% o último idioma é o principal do documento
]{article}

\usepackage{arxiv}
\usepackage[utf8]{inputenc} % allow utf-8 input
\usepackage[T1]{fontenc}    % use 8-bit T1 fonts
\usepackage{hyperref}       % hyperlinks
\usepackage{url}            % simple URL typesetting
\usepackage{booktabs}       % professional-quality tables
\usepackage{amsfonts}       % blackboard math symbols
\usepackage{nicefrac}       % compact symbols for 1/2, etc.
\usepackage{microtype}      % microtypography
\usepackage{lipsum}		% Can be removed after putting your text content
\usepackage{graphicx}
\usepackage{amssymb}
\usepackage{amstext}
\usepackage{amsmath}
\usepackage{amsthm}
\usepackage{subfig}  
\usepackage{stfloats}   
\usepackage{textcomp}  
\usepackage{cite}

\newcommand{\rvx}{\textbf{\emph{x}}}

\newcommand{\rvy}{\textbf{\emph{y}}}

\newcommand{\rvw}{\textbf{\emph{w}}}
\newcommand{\rvW}{\textbf{\emph{W}}}
\newcommand{\rvb}{\textbf{\emph{b}}}
\newcommand{\rva}{\textbf{\emph{a}}}
\newcommand{\rve}{\textbf{\emph{e}}}
\newcommand{\rvh}{\textbf{\emph{h}}}
\newcommand{\rvB}{\textbf{\emph{B}}}

\newtheorem{defi}{Definition}[subsection]

\title{State-of-Charge Estimation of a Li-Ion Battery using Deep Forward Neural Networks}

\author{ \href{https://orcid.org/0000-0003-3557-1957}{\includegraphics[scale=0.06]{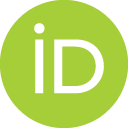}\hspace{1mm}Alexandre Barbosa de Lima} \\
	Polytechnic School of the University of S\~ao Paulo\\
	Department of Energy and Automation\\
	\texttt{alexandreblima@usp.br} 
	\And
	{\hspace{1mm}Maur\''cio B. C. Salles} \\
	Polytechnic School of the University of S\~ao Paulo\\
	Department of Energy and Automation  \\
	\texttt{mausalles@usp.br} \\
	\And
	{\hspace{1mm}Jos\'e Roberto Cardoso} \\
	Polytechnic School of the University of S\~ao Paulo\\
	Department of Energy and Automation  \\
	\texttt{jose.cardoso@usp.br} \\
}

\hypersetup{
pdftitle={A template for the arxiv style},
pdfsubject={q-bio.NC, q-bio.QM},
pdfauthor={David S.~Hippocampus, Elias D.~Striatum},
pdfkeywords={First keyword, Second keyword, More},
}

\begin{document}
\maketitle

\begin{abstract}
This article presents two Deep Forward Networks with two and four hidden layers, respectively, that model the drive cycle of a Panasonic 18650PF lithium-ion (Li-ion) battery at a given temperature using the K-fold cross-validation method, in order to estimate the State of Charge (SOC) of the cell.  The drive cycle power profile is calculated for an electric truck with a 35kWh battery pack scaled for a single 18650PF cell. We propose a machine learning workflow which is able to fight overfitting when developing deep learning models for SOC estimation. The contribution of this work is to present a methodology of building a Deep Forward Network for a lithium-ion battery and its performance assessment, which follows the best practices in machine learning.
\end{abstract}

% keywords can be removed

\section{Introduction}\label{sec:intro}

\subsection{State of the Art and Trends in Li-ion Battery Estimation}

Energy storage acts as a mediator between variable loads and variable sources. Electricity storage is not new. Volta invented the modern battery in 1799. Batteries were implemented in telegraph networks in 1836 \cite{sandia}. The Rocky River Hydroelectric Power Plant in New Milford, Connecticut, was the first major electrical energy storage (EES) system project in the United States. The plant used hydroelectric storage technology through Pumped Hydroelectric Storage (PHS) pumping.

This research is motivated by the study of the application of EES systems in the area of sustainable energy sources.

Hannan et al. \cite{hannan2016} present a detailed taxonomy of the types of energy storage systems taking into account the form of energy storage and construction materials: mechanical, electrochemical (rechargeable and flow batteries), chemical, electrical (ultracapacitor or superconducting magnetic coil), thermal and hybrid.

Recently, industry and academia have given great importance to the electrification of the transport system, given the need to reduce the emission of greenhouse gases. Hybrid electric vehicles, such as the Toyota Prius, or fully electric vehicles, such as the various Tesla models, the Nissan Leaf and the GM Volt, are successful cases in the United States \cite{whit2012}.

The advancement of EES technologies enabled the emergence of the iPod, smartphones and tablets with lithium-ion (li-ion) batteries. If renewable sources, such as solar and wind, become prevalent, the EES will be one of the critical components of the electricity grid, given the intermittent nature of these energy sources \cite{whit2012, luo2015}. EES systems are necessary even when renewable sources are connected to the grid, because it is necessary to smooth the energy supply. For example, the EES of a building or factory can be charged during hours of reduced demand and supply/supplement energy demand during peak hours.

EES technology consists of the process of converting a form of energy (almost always electrical) to a form of storable energy, which can be converted into electrical energy when necessary. EES has the following functions: to assist in meeting the maximum electrical load demands, to provide time-varying energy management, to relieve the intermittency of renewable energy generation, to improve energy quality/reliability, to serve remote loads and vehicles, to support the realization of smart grids, improve the management of distributed/standby power generation and reduce the import of electricity during peak demand periods \cite{luo2015, byrne2017}.

An EES (which can connect to the network or operate in stand-alone mode) consists of two main subsystems: i) storage and ii) power electronics. Such subsystems are complemented by other components that include monitoring and control systems \cite{sandia}.

Lithium-ion battery technology has attracted the attention of industry and academia for the past decade. This is mainly due to the fact that lithium-ion batteries offer more energy, higher power density, higher efficiency and lower self-discharge rate than other battery technologies such as NiCd, NiMH, etc. \cite{motapon2017}.

The efficient use of the lithium-ion battery requires the supervision of a Battery Management System (BMS), as it is necessary that the battery operates under appropriate conditions of temperature and charge (State-Of-Charge (SOC)) \cite{huria2012}. The cell temperature produces deleterious effects on the open circuit voltage, internal resistance and available capacity and can also lead to a rapid degradation of the battery if it operates above a given temperature threshold. Therefore, the modeling of the battery is of paramount importance, since it will be used by the BMS to manage the operation of the battery \cite{motapon2017}.

There are two methods of battery modeling: i) model-driven and ii) data-driven (based on data that is collected from the device) \cite{ren2018}.

Electrothermal models, which belong to the category of model-driven methods, are commonly classified as: i) electrochemical or ii) based on Equivalent Circuit Models (ECM) \cite{motapon2017, huria2012}.

Electrochemical models are based on partial differential equations \cite{jeon2014} and are able to represent thermal effects more accurately than ECM 
\cite{li2014}. However, the first class of models requires detailed knowledge of proprietary parameters of the battery manufacturer: cell area, electrode porosity, material density, electrolyte characteristics, thermal conductivity, etc. This difficulty can be eliminated by characterizing the battery using a thermal camera and thermocouples. But this solution is expensive, time consuming and introduces other challenges such as the implementation of dry air purge systems, ventilation, security, air and water supply, etc. Electrochemical models demand the use of intensive computing systems \cite{huria2012}.

On the other hand, the ECM-based approach has been used for computational/numerical analysis of batteries \cite{huria2012}. In this case, the objective is to develop an electrical model that represents the electrochemical phenomenon existing in the cell. The level of complexity of the model is the result of a compromise between precision and computational effort. Note that an extremely complex and accurate ECM may be unsuitable for application in embedded systems.

The most recent literature show that the machine learning approach, based on deep learning algorithms  is the state of the art in the area \cite{ren2018, zhao2017, chemali2018, chen2019, goodf2016, lecun2015, mallat2016, shervin2019, silver2018, lstm-li-ion-2018, Kollmeyer2017}. Machine learning is a branch of AI, as will be seen in section \ref{sec:dl}.

Chemali et al \cite{chemali2018} compared the performance of Deep Neural Networks (DNN) with those of other relevant algorithms that have been proposed since the second half of the 2000s. The article shows that the SOC estimation error obtained with deeep learning  is less than the following methods:

\begin{itemize}
	\item Model Adaptive-Improved Extended Kalman Filter (EKF) \cite{sepasi2014};
	\item Adaptive EKF with Neural Networks \cite{chark2010};
	\item Adaptive Unscented Kalman Filter (AUKF) with Extreme Machine Learning \cite{du2014};
	\item Fuzzy NN with Genetic Algorithm \cite{lee2014}; and
	\item Radial Bias Function Neural Network \cite{chang2013}.   
\end{itemize}

%In addition, Chemali et al \cite{chemali2018} show that DNN allow the behavior of a lithium ion cell at different temperatures to be modeled by the weights of a single neural network, which can accurately and robustly estimate the SOC in a given temperature range (eg $0 ^ o \, \text {C} \, \text {a} \, 40 ^ o \text {C} $). Note that the ECM strategy does not have this flexibility. The ECM of a cell at $20^o$ C is different from the ECM of the same cell at $40^o$ C.

%The article by Ren et al \cite{ren2018} shows that advances in deep learning have introduced new data driven approaches for estimating the Remaining Useful Life (RUL) of lithium ion batteries. Ren et al \cite{ren2018} used the NASA Ames Research Center training examples database \cite{saha2007}. The results of Ren et al \cite{ren2018} suggest that the RUL estimation approach using deep learning is superior to that of other machine learning methods, such as Bayesian regression and Support Vector Machines (SVM).

Estimating the SOC of lithium ion cells in a BMS by means of deep learning offers at least two significant advantages over model driven approaches, namely: i) neural networks are able to estimate the non linear functional dependence that exists between voltage, current and temperature (observable quantities) and unobservable quantities, such as SOC, with great precision  and ii) the problem of identifying ECM parameters is avoided.

\subsection{Contribution of the paper}

In relation to the literature already mentioned, this paper takes a step back as far as reasons related to the methodology adopted for the construction and performance measurement of a deep learning model are concerned. %for the estimation of SOC (and other parameters) of Li-Ion batteries. 

First, we work with Deep Feedforward Networks (DFN) as a baseline family model, as they form the basis of many important commercial applications \cite{goodf2016}. Our preliminary results show that a simple architecture with four hidden layers is already quite interesting. 
We do not work with Recurrent Neural Networks (RNN)  because they are suitable for sequential data processing  problems such as machine translation.

Second, the central challenge in machine learning is that our model has to perform well on new, unseen inputs, not just those on which our algorithm was trained. This ability is called generalization. What separates machine learning from traditional optimization is that we want the generalization error, or test error, to be as low as possible \cite{goodf2016}. To do this, we need a test set. In this work we followed the best practice of breaking the test set into two separate sets: validation and test sets \cite{cholletPy}. That is, the generalization power of the models were measured against validation and test sets. The validation set is used to fine tune the network hyperparameters. % during the K-fold cross-validation phase. 

\begin{figure}[h]
	\begin{center}
		\includegraphics*[scale = 0.45]{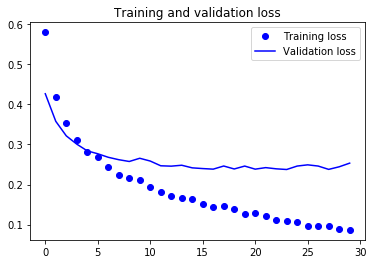}
	\end{center}
	\caption{Training and generalization errors behave differently. The horizontal axis represents the number of epochs. The vertical axis is the loss function. Note that the model starts to overfit around the fifth epoch.}
	\label{fig:training_validation_loss_cats_dogs_VGG16}
\end{figure}

Third, as a corollary of generalizatin, the central problem in machine learning, namely overfitting, has not been properly addressed, to the best of our knowledge, in the recent mainstream literature of SOC estimation of Li-ion batteries using deep learning. 
Thus, we have to apply concepts from statistical learning theory \cite{goodf2016}. Overfitting occurs when the gap between the training error and generalization error is too large. The processing of mitigating overfitting  is called regularization, which can be defined as any modification we make to a learning algorithm with the goal of reducing its generalization error but not its training error.  To see this phenomenon, one has to plot the training and generalization learning curves, see  Fig. \ref{fig:training_validation_loss_cats_dogs_VGG16} for instance.

Note  that the literature of machine learning presents a solid framework for solving deep learning problems, such as model evaluation and the attack on overfitting, among others \cite{goodf2016, cholletPy, kevin2012}. The satisfactory result obtained in this article in terms of a low generalization error takes overfitting into account.

Fourth, as mentioned before, we have to consider the optimization problem in the context of deep learning, which completely differs from traditional optimization algorithms in several ways. In this work, we use algorithms with adaptive learning rules, such as RMSProp and Adam, which include the concept of momentum, allowing faster convergence than Stochastic Gradient Descent (SGD), at the cost of more computation. The learning rate is one of the most difficult hyperparameters of a artificial neural network (ANN) to be configured as it significantly affects the model performance. Note that small values of the learning rate result in a slow convergence of deep learning. On the other hand, if the learning rate is too large, gradient descent can overshoot the minimum. It may fail to converge, or even diverge. %Our simulations used the Adam optimizer, which presented slightly better results than RMSProp.

\begin{figure}[h]
	\begin{center}
		\includegraphics*[scale = 0.35]{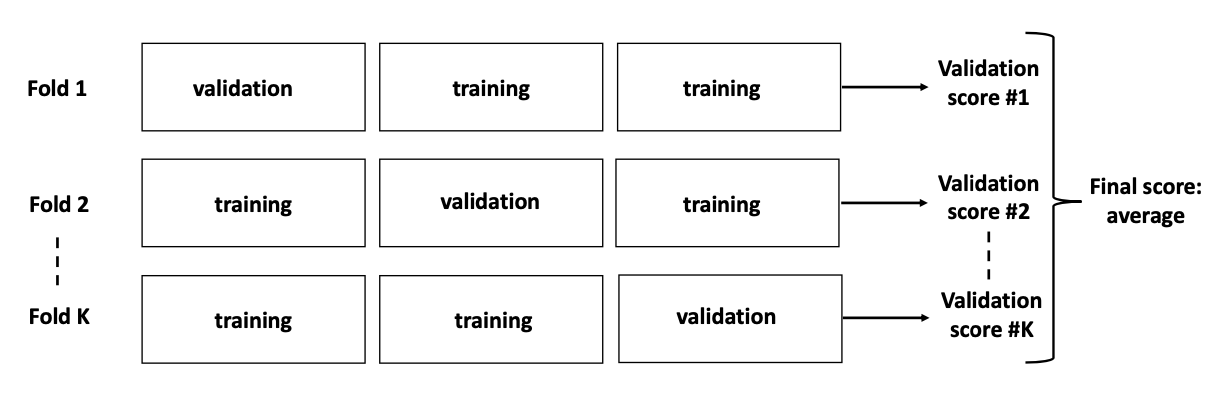}
	\end{center}
	\caption{K-fold cross-validation.}
	\label{fig:K-fold-cross-valid}
\end{figure}

Fifth, the validation error is estimated by taking the average validation error across $K$ trials. 
We use a simple, but popular solution, called $K$-fold cross-validaton (Fig. \ref{fig:K-fold-cross-valid}), which consists of splitting the available training data into two partitions (training and validation), instantiating $K$ identical models, for each fold $k\in \{1,2,\ldots,K\}$, and training each one on the training partitions, while evaluating on the validation partition. The validation score for the model used is then the average of the $K$ validation scores obtained. This procedure allows network hyperparameters to be adjusted so that overfitting is mitigated \cite[p. 23]{kevin2012}. It is usual to use about $80\%$ of the data for the training set, and $20\%$ for the validation set. Note that the validation scores may have a high variance with regard to the validation split. Therefore, $K$-fold cross-validaton help us improve the reliability when evaluating the generalization power of the model.

%Seventh, deep learning models tend to be good at fitting to the training data, but the real challenge is generalization, not fitting. To do this, you need a test set, which has data that the deep learning has never seen before. In this work, the generalization power of the models were measured against a test set, after the tuning of the network hyperparameters during the K-fold cross-validation phase.

\subsection{Organization of the paper}
The remainder of the paper is organized as follows.  Section \ref{sec:dl} presents an overview of AI, machine learning and deep learning for the reader who is not familiar with the subject. Section \ref{sec:results} presents our experimental results. Finally, section \ref{sec:conclusions} presents our conclusions.

\section{Artificial Intelligence and Deep Learning}\label{sec:dl}

\subsection{The Notion of Neural Nets}\label{subsec:nn}

\textit{The Handbook of Artificial Intelligence} presents the following operational definition\footnote{There is no consensus on the concept of intelligence.} for Artificial Intelligence (AI) \cite{handbook-AI-vol1}:

\begin{quotation}
	``Artificial Intelligence (AI) is the part of computer science concerned with designing intelligent computer systems, that is, systems that exhibit the characteristics we associate with intelligence in human behavior - understanding language, learning, reasoning, solving problems, and so on''. 
\end{quotation}

There are two main lines of research in AI: the connectionist and the symbolic. According to Boden \cite{boden2016}, both were inspired by the seminal article entitled  \textit{A Logical Calculus of the Ideas Immanent in Nervous Activity} (1943), by Warren Sturgis McCulloch and Walter Pitts \cite{pitts1943}, the first modern computational theory\footnote{The theory is considered modern because it employs the mathematical notion of computation established by Turing in 1936 \cite{turing1936}.}. The literature recognizes that the research carried out by McCulloch and Pitts is the pioneering work in AI \cite{russel}.

\begin{figure}[ht]
	\begin{center}
		\includegraphics[scale = 0.5]{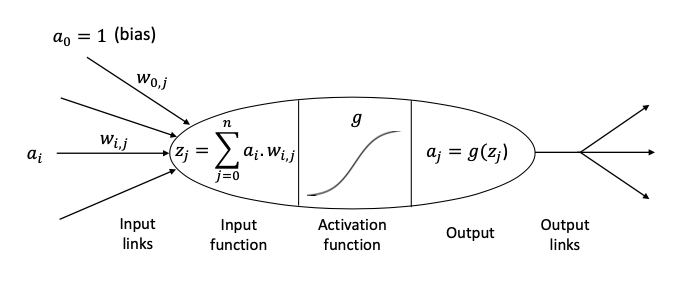}
	\end{center}
	\caption{McCulloch and Pitts model}
	\label{fig:neuron-unit}
\end{figure}

Fig. \ref{fig:neuron-unit} illustrates the McCulloch and Pitts artificial neuron model.
%A Fig. \ref{fig:neuron-unit} (fonte: \cite{russel}) ilustra o modelo de neurônio artificial  McCulloch e Pitts \cite{pitts1943}. 

The artificial neuron (or unit) of Fig. \ref{fig:neuron-unit} has the following characteristics:
%O neurônio (ou unidade)  artificial da Fig. \ref{fig:neuron-unit} possui as seguintes características:

\begin{itemize}
	
	\item input signals (activations) ($a_1, a_2, \ldots, a_n $) are $0$ or $1$ bits. The external signal $a_0 = 1$ is known as the bias\footnote{The bias plays the role of the intercept $b$ in the simple linear regression model $y = wx + b$.};
	%sinais (ativações) de entrada ($a_1, a_2, \ldots, a_n$) são  bits $0$ ou $1$. O sinal externo  $a_0 = 1$ é o \textit{bias};
	
	\item synaptic weights of the j-th neuron: $w_0, w_1, \ldots, w_n$; and%pesos sinápticos do j-ésimo neurônio: $w_0, w_1, \ldots, w_n$; e
	
	\item $a_j$ denotes the output signal (or output activation) of the jth neuron, given by%$a_j$ denota o sinal de saída (ou ativação de saída) do j-ésimo neurônio, dado por:
	
	\begin{equation}
	\label{eq:z_j}
		z_j  = \sum_{i=0}^{n} w_{i,j}a_i 
	\end{equation} 
	
	\begin{equation}
	\label{eq:a_j}
		a_j  = g(z_j) = \{0,1\} 
	\end{equation} 
	
where $z=f(a)$ is the input function and $g (z)$ is a nonlinear activation function. The output is binary (bit $ 0 $ or bit $ 1 $); therefore, the McCulloch and Pitts model is said to have the `` all or nothing '' property.%em que $g(z)$ é uma função de ativação não linear. A saída é binária (bit $0$ ou bit $1$);  por conseguinte, diz-se que o modelo de McCulloch e Pitts tem a propriedade ``tudo ou nada''.  	
\end{itemize}

The activation function of the McCulloch and Pitts model is the Heaviside function (unit step function)

\begin{equation}
\label{eq:degrau}
g(z)= \left\{
\begin{array}{ll}
1, & z  \geq 0	\\
0, & z<0
\end{array}
\right.
% CUIDADO: Não esquecer de colocar "." depois do "\right" !!
\end{equation}

%which is shown by Fig. \ref{fig:degrau}).
$g(z)=1$ se $z  \geq 0	$ ou $g(z)=0$ para $z<0$.

\begin{figure}[h]
	\begin{center}
		\includegraphics*[scale = 0.30]{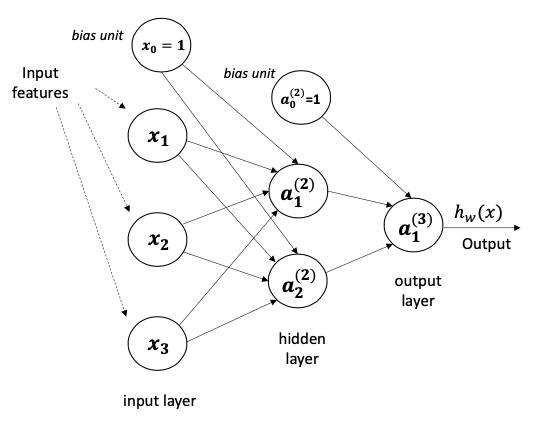}
	\end{center}
	\caption{Example of a densely connected neural net with one hidden layer. The hidden layer has three units ($a_0^{(2)}$, $a_1^{(2)}$, and $a_2^{(2)}$). The input features are the input signals $\rvx$. The function $\rvy=\rvh_{\rvw}(\rvx)$ is called the model or hypothesis.}
	\label{fig:ann}
\end{figure}

We take the opportunity to make a necessary digression on the neuron model of Fig. \ref{fig:neuron-unit}, in order to present the intuition behind the fact that modern ANN (see Fig. \ref{fig:ann}) are able to approximate nonlinear functions with arbitrary precision, at least in theory (universal approximation theorem \cite{cybenko1989}). The explanation below considers a network with only one layer of  $N + 1 = M$ neurons in parallel, where each neuron is excited by the input signal $\rva = \{a_0, a_1, \ldots, a_N \}$ , not necessarily binary. This layer is known as the hidden layer.

%Aproveitamos a oportunidade para fazer uma pequena digressão sobre o modelo de neurônio que é usado atualmente e também para apresentar a intuição por trás do fato de as redes neurais conseguirem aproximar funções não lineares com precisão arbitrária, pelo menos em teoria (teorema de aproximação universal \cite{cybenko1989}). A exposição a seguir considera uma rede com apenas uma camada de $N+1=M$ neurônios em paralelo, em que cada neurônio é excitado pelo sinal de entrada $\rva = \{a_0, a_1, \ldots, a_N\}$, não necessariamente binário. 

Let's rewrite (\ref{eq:z_j}) in a vectorized form
%Vamos rescrever (\ref{eq:z_j}) na forma vetorizada

\begin{equation}
\label{eq1:z_j}
	z_i  = \rvW^T \rva 
\end{equation} 

%em que adotamos a notação $\rvW$ para o vetor de pesos sinápticos, $\rva$ para o vetor de entradas  e $T$ denota a transposição de matrizes\footnote{Este artigo assume que os vetores são do tipo coluna, como é usual na literatura de processamento de sinais \cite{ali2003}.} . 
where we adopt the notation $\rvW $ for the vector of synaptic weights, $\rva$ for the vector of entries and $T$ denotes the transposition of matrices \footnote{This article assumes that vectors are always column vectors, as it is usual in the signal processing area.}.

Consider the Discrete Fourier Transform (DFT) \footnote{The imaginary unit is represented by  $j$.} of an input signal $\rvb = \{b_0, b_1,  \ldots, b_N\}$ given by 
\begin{equation}
\label{eq:dft}
B[k]= \left\{
\begin{array}{ll}
\overset{N}{\underset{i=0}\sum} (e^{-j 2\pi \frac{k}{M}i}) b_i, & 0\leq k \leq N\\
0, & \text{otherwise}
\end{array}
\right.
% CUIDADO: Não esquecer de colocar "." depois do "\right" !!
\end{equation}
and the corresponding inverse transformation, called Inverse Discrete Fourier Transform (IDFT) \cite{opp2009}
\begin{equation}
\label{eq:idft}
z_i= \left\{
\begin{array}{ll}
\overset{N}{\underset{k=0}\sum} (\frac{1}{M} e^{j2\pi\frac{k}{M} i}) B[k], & 0\leq k\leq N\\
0, & \text{otherwise}
\end{array}
\right.
% CUIDADO: Não esquecer de colocar "." depois do "\right" !!
\end{equation}

%Reescrevendo (\ref{eq:idft})  na forma vetorizada, obtemos
Rewriting (\ref{eq:idft}) in vectorized form, we obtain

\begin{equation}
\label{eq1:idft}
	z_i  = \rvW^T \rvB
\end{equation} 

where $\rvW=\{(\frac{1}{M} e^{j2\pi\frac{k}{M} i})\}, 0\leq k\leq N$, and $\rvB=\{B_0, B_1\ldots\,B_N\}$.

Compare (\ref{eq1:idft}) and (\ref{eq1:z_j}). Note that these equations are equal if $\rva = \rvB$. Therefore, Eq. (\ref{eq1:idft}) suggests that it would be possible to represent the function $z = f(a)$ through a neural network that uses the weights $\{\frac{1}{M} e^{j2 \pi\frac{k}{M} i}\}, 0 \leq k \leq N $. Eq. (\ref{eq1:idft}) looks like a Fourier series. 

Remember that Fourier showed in 1807 that an arbitrary and aperiodic function $f (t)$ defined in a finite interval $T_0$ can be reconstructed from a trigonometric series called the Fourier series \cite{lathi1998}. So ``there is nothing new under the sun''. Electrical engineers are well familiarized with this notion.

As mentioned before, the universal approximation theorem states that a feedforward network with a linear input and at least one hidden layer of artificial units with an non linear activation function can approximate any ``function''\footnote{In fact, any Borel function. This concept is beyond the scope of this paper and will not be discussed.} from one finite-dimensional space to another with any desired nonzero amount of error, provided that the network is given enough units \cite{goodf2016, cybenko1989}. However, the theorem does not say what the number of units in the hidden layer should be. We also have no guarantees that the training algorithm will be able to learn that function. This may be due to the existence of local minimums in the cost function to be optimized.

\begin{figure}[h]
	\begin{center}
		\includegraphics*[scale = 0.20]{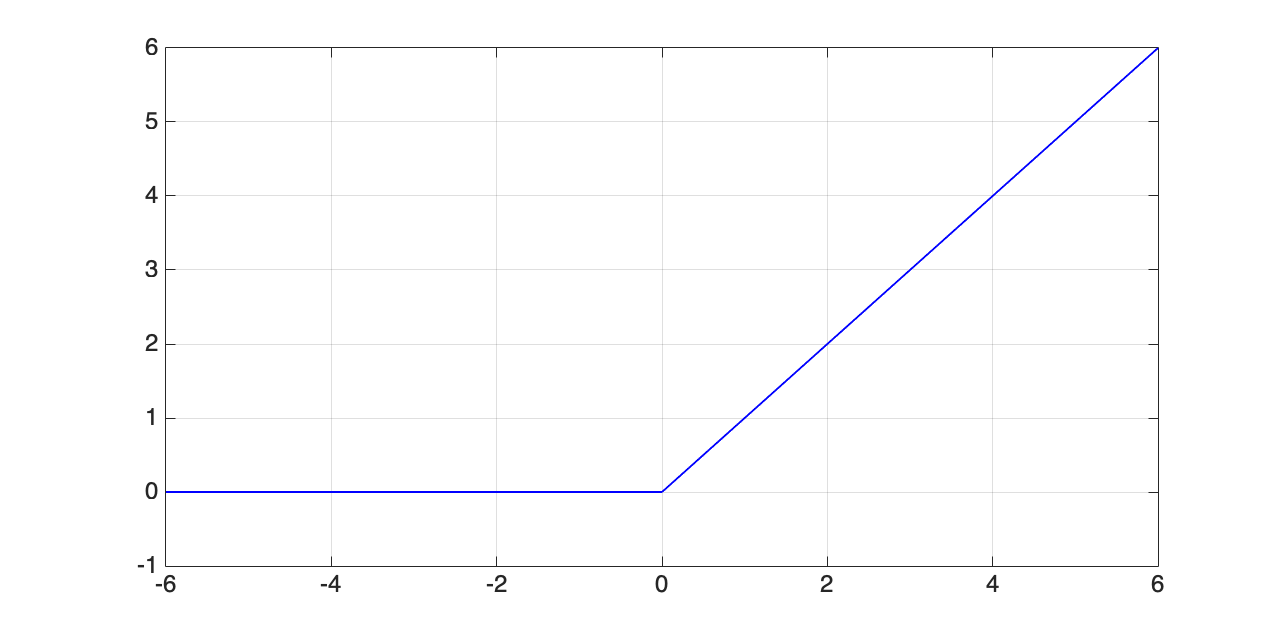}
	\end{center}
	\caption{Relu activation funcion.}
	\label{fig:relu}
\end{figure}

Nowadays, the activation function called REctified Linear Unit (Relu) (see Fig\ref{fig:relu}), given by

\begin{equation}
\label{eq:relu}
	g(z)  = \text{max}\{0,z\}
\end{equation} 

is commonly used in ANN \cite{goodf2016}. 

An ANN is a distributed parallel processing system, inspired by the processing structure of the human brain. The ANN technique is a form of non-algorithmic computation of functions.	

The connectionist approach uses ANN. The symbolic line, sometimes called the symbolic AI (``Good Old Fashioned AI (GOFAI) \cite{boden2016}) follows the logical tradition and had John McCarthy and Allen Newell, among others, as some of its great exponents 
\cite{ newell1980}.

Since 2006, the connectionist line has gained prominence,  due in large part to the fundamental contributions of the researchers Yoshua Bengio, Geoffrey Hinton, and Yann LeCun, who were awarded the 2018 A. M. Turing Award \cite{acm2018}. Nowadays AI research is  dominated by systems that use ANN, also known as Deep Learning \cite{goodf2016}. Fig. \ref{fig:DL-6layers} shows a deep learning model for handwritten digit recognition (a classical problem in computer vision), first solved by LeCun et al \cite{Cun90handwrittendigit}. The model has one input layer, four hidden layers, and one output layer.

\begin{figure}[h]
	\begin{center}
		\includegraphics*[scale = 0.35]{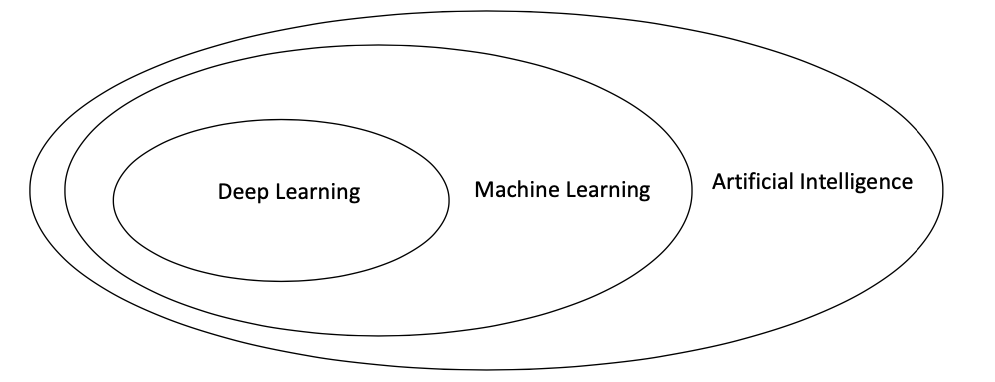}
	\end{center}
	\caption{Relationship between AI, machine learning and deep learning.}
	\label{fig:venn}
\end{figure}

\begin{figure}[h]
	\begin{center}
		\includegraphics*[scale = 0.35]{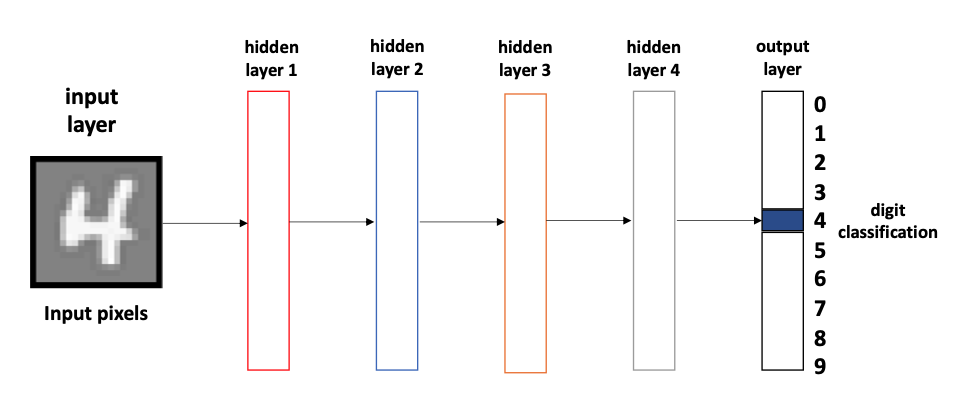}
	\end{center}
	\caption{Handwritten Digit Recognition with a 6 layers ANN.}
	\label{fig:DL-6layers}
\end{figure}

\subsection{Machine Learning}\label{subsec:ML}

Starting at 1952, Arthur Samuel, the pioneer of machine learning in the era of digital computers \cite{McCarthy1990}, wrote a series of programs for IBM computers that learned to play checkers \footnote {The work was developed during Samuel's free time.} through a learning process that was not based on neural networks. The program was shown on TV at 1956, making a strong impression on public opinion \cite{samuel1959, ertel2017}. 
%Samuel proposed the following definition for machine learning in 1959 \cite{Ng2019}:

%\begin{quote}
%	``Machine Learning is the area of knowledge that gives computers the ability to learn without being explicitly programmed.''
%\end{quote}

%In other words, a machine learning system is trained for a task, instead of being explicitly programmed for it.

Tom Mitchell gives a definition of machine learning \cite[p. 2]{tom1997}

\begin{quote}
`` A computer  is said to \textbf{learn} from experience \textit{E} with respect to some class of tasks \textit{T} and performance measure \textit{P}, if its performance at tasks in  \textit{T}, as measured by \textit{P}, improves with experience \textit{E}''.
\end{quote}

As Samuel have shown, checkers is a problem that can be solved using machine learning algorithms. According to Mitchell's definition, we have that:

\begin {itemize}
	\item \textit{E} = `` training '' of the program, which consists of the program playing a sufficiently large number \footnote{For example, hundreds of thousands of games.} of checkers games against itself.
	\item \textit{T} = playing games.
	\item \textit{P} = percent of games won against opponents.
\end {itemize}

Machine learning  algorithms can be divided into four groups \cite{cholletPy}:

\begin{itemize}

	\item Supervised Learning: in which the man provides the input data (training set), as well as the expected responses from the data (a label is attached to each data ). The output of the processing is a set of rules, which will be applied on new input data, in order to produce original responses (inferences of the classification or regression type).

	\item Unsupervised Learning: for training a descriptive model capable of recognizing patterns. There is no teacher. Applications: discover shopping patterns in supermarkets and data clustering tasks (eg, identifying groups of people who share common interests such as sports, religion or music).

	\item Self-supervised Learning: is a type of supervised learning; there are still labels involved (because learning needs to be supervised by something), but they are generated from the input data, usually using a heuristic algorithm.

	\item Reinforcement Learning: a program is trained to interact in a environment through actions to achieve a goal. Learning uses a rewards and/or punishment mechanism. The AlphaZero algorithm \footnote{AlphaZero defeated the world champion of Go, Lee Sedol, in a challenge sponsored by Google in 2016 in South Korea.}, developed by Google DeepMind, uses this technique \cite{alphazero}.

\end{itemize}

Supervised learning, which is used in this work, involves the training of a model or hypothesis. To this end, given a set of training data (or training set), the learning algorithm aims to fit a model that minimizes a metric of choice.

In 1959, Bernard Widrow and Martian E. Hoff discovered the famous machine learning algorithm called Least Mean Square (LMS) \cite {widrow1960} The LMS algorithm belongs to the family of stochastic gradient algorithms and uses the method of optimization of the gradient descent, which is also used in deep learning, whose objective is to find the values of the parameters that minimize a given function, called objective function or cost function.

Adaline implements supervised machine learning, since the desired response for each input pattern is provided.

In 1958, Frank Rosenblatt  presented the theory of a hypothetical nervous system called the Perceptron \cite{rosenblatt1958}. Perceptron is a single-layer ANN with a learning rule capable of implementing a linear classifier.

A Perceptron follows the feed-forward processing model, from left (inputs) to right (output (s)). %Fig. \ref{fig:perceptron}(source: \cite{russel}) shows: (a) perceptron with two inputs and two output units. (b) neural network with two inputs, a hidden layer with two neurons and two output units.

The 1980s were marked by the discovery of the backpropagation algorithm,  which updates the gradient for multilayer networks \cite {hinton1986}. The wide dissemination of the results in the collection \textit{Parallel Distributed Processing} \cite{hinton1986}, edited by Rumelhart and McClelland, caused great excitement in the areas of Computer Science and Psychology.

In fact, the backpropagation algorithm was discovered independently several times \cite{bryson, werbos74, parker85, lecun1986}. 

\subsection{Statistical Learning}\label{subsec:prob-ML}

This section introduces a very short review of the main concepts about stochastic processes (also called data-generating processes by the literature of deep learning \cite{goodf2016}) and statistical learning that are used in this paper. The purpose is just to indicate what is important to be known without entering into the details. The interested reader should refer to the appropriate literature \cite{kevin2012, papoulis91}.

\begin{defi}[Stochastic Process]\index{stochastic processes} 
	\label{def:PE}
	Let  $T$ be an arbitrary set. A stochastic process is a family $\{\rvx_t, t \in T\}$, such that, for each $t \in T$, $\rvx_t$ is a random variable.  $\quad \blacksquare$
\end{defi}

When the set $T$ is the set of integer numbers $\mathbb{Z}$, then  $\{\rvx_t\}$ is a discrete time stochastic process (or random sequence);  $\{\rvx_t\}$ is a continuous time stochastic process if $T$ is taken as the set of real numbers $\mathbb{R}$. 

The random variable $\rvx_t$ is, in fact, a function of two arguments $\rvx(t,\zeta),\, t\in T,\, \zeta \in \Omega$, given that it is defined over the sample space $\Omega$.  For each $\zeta \in \Omega$ we have a realization, trajectory or time series $x_t$. The set of all realizations is called ensemble. Each trajectory  is a function or a non-random sequence and for each fixed $t$, $x_t$ is a number.

A process $\rvx_t$ is completely specified by its \textit{finite-dimensional distributions} or $n$-order probability distribution functions\index{probability distribution function}, as:

\begin{equation}
	\label{eq:dist-PE}
	F_{\rvx}(x_1,x_2,\ldots,x_n;t_1,t_2, \ldots,t_n)  = 
	P\{\rvx(t_1)\leq x_1, \rvx(t_2)\leq x_2, \ldots, \rvx(t_n)\leq x_n\}
\end{equation}
in which  $t_1,t_2, \ldots,t_n$ are any elements of $T$ and $n\geq1$. 

The first order probability distribution function is also known as Cumulative Distribution Function - CDF\index{Cumulative Distribution Function - CDF}.	

The  \textit{probability density function - PDF}\index{probability density function - PDF} is given by:

\begin{equation}
	\label{eq:pdf-n}
	f_{\rvx}(x_1,x_2,\ldots,x_n;t_1,t_2, \ldots,t_n)  
	= \frac{\partial^n F_{\rvx}(x_1,x_2,\ldots,x_n;t_1,t_2, \ldots,t_n)}{\partial x_1 \partial x_2 \ldots \partial x_n}.
\end{equation}	

Applying the conditional probability density formula, 
\begin{equation}
	\label{eq:pdf-cond}
	f_{\rvx}(x_k | x_{k-1},\ldots,x_1) = \frac{f_{\rvx}(x_1,\ldots,x_{k-1},x_k)}{f_{\rvx}(x_1,\ldots,x_{k-1})},
\end{equation}
in which $f_{\rvx}(x_1,\ldots,x_{k-1},x_k)$ denotes $f_{\rvx}(x_1,\ldots,x_{k-1},x_k;t_1,\ldots,t_{k-1},t_k)$, repeatedly over $f_{\rvx}(x_1,\ldots,x_{n-1},x_n)$ we get the probability chain rule
\begin{equation}
	\label{eq:pdf-chain-rule}
	f_{\rvx}(x_1,x_2,\ldots,x_n)  
	= f_{\rvx}(x_1)f_{\rvx}(x_2|x_1)f_{\rvx}(x_3|x_2,x_1)\ldots f_{\rvx}(x_n | x_{n-1},\ldots,x_1).
\end{equation}

When $\rvx_t$ is a sequence of \textit{mutually independent} random variables, (\ref{eq:pdf-chain-rule}) can be rewritten as
\begin{equation}
	\label{eq:pdf-vas-mut-ind}
	f_{\rvx}(x_1,x_2,\ldots,x_n)  
	= f_{\rvx}(x_1)f_{\rvx}(x_2)\ldots f_{\rvx}(x_n).
\end{equation}

\begin{defi}[Purely Stochastic Process]\index{Purely Stochastic Process}
	\label{def:PPE}
	A \textit{purely stochastic} process $\{\rvx_t, t \in \mathbb{Z}\}$ is a sequence of mutually independent random variables.  $\quad \blacksquare$
\end{defi}

\begin{defi}[IID Process]\index{IID Process}
	\label{def:PIID}
	\begin{sloppy}
		An \textit{Independent and Identically Distributed (IID)} process $\{\rvx_t, t \in \mathbb{Z}\}$, denoted by $\rvx_t \sim \text{IID}$, is a purely stochastic and identically distributed process.  $\quad \blacksquare$
	\end{sloppy}
\end{defi} 

% Como já foi dito anteriormente, o desafio central em machine learning é que o algoritmo tenha um bom desempenho sobre o conjunto de teste. Os conjunto de treinamento e de teste são gerados pelo mesmo processo aleatório. Tipicamente, assume-se que os exemplos em cada conjunto são independentes e que os conjuntos de treinamento e de teste sejam identicamente distribuídos. Estas hipóteses são conhecidas coletivamente como premissas i.i.d.

% Contudo, o teorema é válido quando se trabalha com a média de todos os os possíveis data-generating processes. Ora, mas isso não ocorre na vida real, que impões restrições ao data-generating processes. Assim, espera-se que, em aplicações práticas, é realista pensar em projetar algoritmos que possuam um bom desempenho para um dado data-generating process. 

% Indo mais longe, o objetivo da pesquisa em algoritmos de machine learning não é procurar peor um algoritmo de aprendizado universal. Em vez disso, temos que entender quais são as características estocásticas do nosso dataset. de forma a projetar, validar e testar um algoritmo que seja eficiente para aquele conjunto de dados.

As mentioned earlier, the central challenge in machine learning is that the algorithm performs well on the test set. The training and test sets are generated by the same data-generating process. Typically, it is assumed that the examples in each set are independent and that the training and test sets are identically distributed. These assumptions are collectively known as IID.

The \textbf{no free lunch theorem} \cite{wolpert1996} says that, considering the average over all the possible data-generating processes, any machine learning algorithm has the same error rate when evaluated on previously unobserved examples. That is, there is not, at least in theory, a machine learning algorithm that is better than all others for all cases.

However, the no free lunch theorem is valid only when working with the average over all the possible data-generating processes. Fortunately, that does not happen in real life, as the physical data-generating process of a Li-Ion cell is a result of the restrictions imposed by the real world over all the possible data-generating processes. Thus, in practical applications, it is realistic to think about designing algorithms that have a good performance for a given Li-Ion cell.

Going further, the goal of research in machine learning is not to look for a universal learning algorithm. Instead, we have to understand what the stochastic characteristics of our dataset are, in order to design, validate and test an algorithm that is efficient for that specific data set.

\subsection{ANN as Deep Learning}\label{subsec:DL}

Interest in ANNs has resurfaced with the advent of the Deep Belief Nets in 2006 \cite{hinton2006}. The work of Hinton, Osindero and Teh demonstrated that a type of DNN could be trained with high efficiency. Their research triggered the current wave of research in ANN, which popularized the term deep learning.
		
Deep learning denotes the idea of a neural network with multiple hidden layers that has the ability to partition the representation of an entry into multiple layers (see Fig.\ref{fig:DL-6layers}). The number of layers in a model corresponds to the depth of the network \cite{cholletPy}.

The success of deep learning today is due to: i) the emergence of Big Data, which made it possible to store data for training in databases with tens of millions of examples, ii) the advent of Graphics Processing Unit (GPU), and iii) advances in algorithms.

\begin{figure}[h]
	\begin{center}
		\includegraphics*[scale = 0.5]{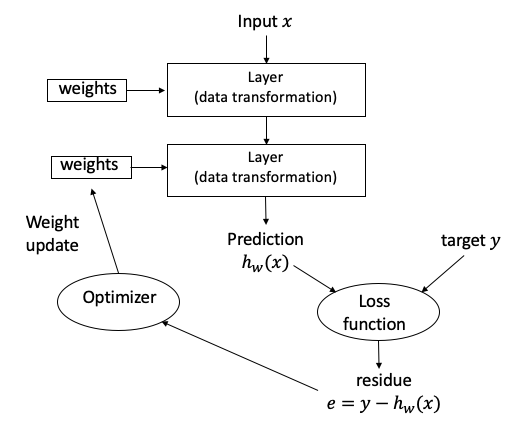}
	\end{center}	
	\caption{How deep learning works.}
	\label{fig:DL}
\end{figure}

Fig. \ref{fig:DL} illustrates how deep learning works \cite{cholletPy}. The variables $\rvx$ and $\rvy$ denote the input signal (training example) and the desired signal (target), respectively. The function $\rvh_{\rvw}(\rvx)$ is the mathematical model or hypothesis. The estimation error (residual or loss score) is given by $\rve = \rvy - \rvh_{\rvw}(\rvx)$. At startup, random values are assigned to the $\rvw$ weights of the network, so the value of the initial residue is high. However, in the course of processing the training examples, the weights are adjusted incrementally in the correct direction; at the same time, the value of the loss function decreases. This is the training loop, which, being repeated enough times, typically dozens of iterations over thousands of examples, produces weights that minimize the cost function. A network is considered trained when the minimum of the cost function is reached.

Deep learning has achieved the following advances, all in historically difficult areas of machine learning, such as \cite{silver2018, cholletPy}:
%O deep learning alcançou os seguintes avanços, todos em áreas historicamente difíceis de machine learning  \cite{cholletPy},  \cite{silver2018}:

\begin{itemize}
	\item nonlinear regression;
	\item superhuman image classification;
	\item voice recognition at an almost human level;
	\item transcription of almost human handwriting;
	\item automatic translation;
	\item text to speech conversion;
	\item autonomous vehicle driving at an almost human level; and
	\item superhuman performance in games like chess, shogi and Go.
\end {itemize}

However, deep learning has limitations. Current ANN architectures do not have the power to track statistical changes of the  training data in real time. In other words, deep learning is not yet adaptive. Note that training a DNN at a dedicated workstation can take days or weeks. This is due to the computational complexity of deep learning. 

Furthermore, the science of deep learning is not like mathematics or physics, in which theoretical advances can be achieved with a chalk and a blackboard. Deep learning is an engineering science \cite{allaire2017}, as it does not yet have a mathematical formalism like that of the area of adaptive filtering. For example, as we have stated before, there is no design criterion for the number of layers in the network, much less for the number of neurons in a hidden layer. The field is driven by experimental discoveries. But of course, there are best practices to be followed \cite{cholletPy}.
%Contudo, o deep learning possui limitações. As arquiteturas atuais de ANN profundas não têm o poder de rastrear alterações nos dados de treinamento em tempo real. Note-se que o treinamento de um rede neural profunda em uma estação de trabalho dedicada pode levar dias ou semanas. E isto se deve à complexidade computacional do deep learning. Dito de outra forma, o deep learning ainda não é adaptativo. 
%Além disso, a ciência do deep learning não é como a Matemática ou a Física, em que avanços teóricos podem ser alcançados com um giz e uma lousa. O deep learning é uma ciência da Engenharia \cite{allaire2017}, pois ainda não possui um formalismo matemático como o da área de filtragem adaptativa. Por exemplo, não há um critério de projeto para o número de camadas na rede, muito menos para o número de neurônios em uma camada oculta. O campo é impulsionado por descobertas experimentais.

\subsection{A Workflow for Approaching Deep Learning Problems}\label{subsec:workflow}

\begin{figure}[h]
	\begin{center}
		\includegraphics*[scale = 0.2]{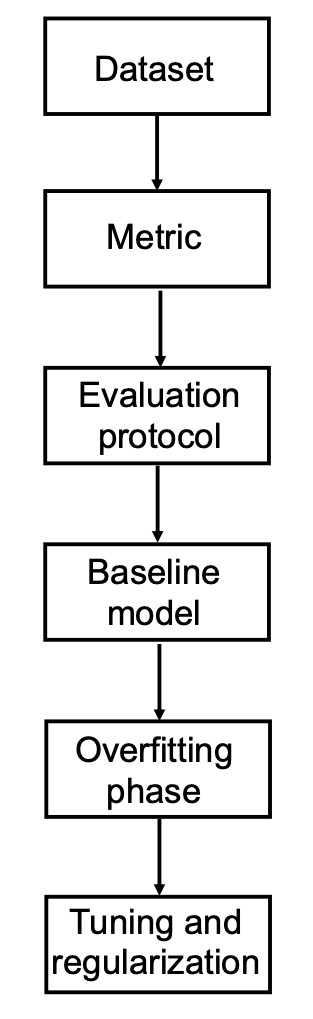}
	\end{center}	
	\caption{Deep learning workflow.}
	\label{fig:workflow-deep-learning}
\end{figure}

In this paper, we follow an adapted version of the supervised machine learning workflow proposed by Chollet (see Fig. \ref{fig:workflow-deep-learning}) \cite{cholletPy}:

\begin{enumerate}
	\item Choose a reliable dataset. If you do not find a dataset, collect the data of interest, and annotate it with labels.
	\item Choose how you will measure success on your problem. Which metrics will you monitor on your validation data?
	\item Determine your evaluation protocol: K-fold cross-validation? Which portion of the data should you use for validation?
	\item Develop a baseline model with statistical power.
	\item Develop a model that overfits.
	\item Regularize your model and tune its hyperparameters, based on performance on the validation data. 
\end{enumerate}

\section{Experimental Results}\label{sec:results}

\subsection{Deep Learning Methodology and Modeling}\label{subsec:metodo}

We selected the  2.9 Ah Panasonic 18650PF Li-ion Battery Data provided by Dr. Phillip Kollmeyer, University of Wisconsin-Madison \cite{Kollmeyer2018}. Note that this dataset has been used by some of the top works in the area \cite{zhao2017, chemali2018, lstm-li-ion-2018, Kollmeyer2017}.

A  series of nine drive cycles were made public, among them a Neural Network (NN) cycle, which is the cycle of our interest. More specifically, the simulations in this section present the results for data collected at a temperature of $25^o$ C.

The NN drive cycle was designed to have some additional dynamics which are useful for training neural networks.  The drive cycle power profile is calculated for an electric Ford F150 truck with a 35kWh battery pack scaled for a single 18650PF cell.

\begin{figure}%
	\centering
	\subfloat[]{
		\label{fig:temperature-SOC}%
		\includegraphics*[scale=0.35]{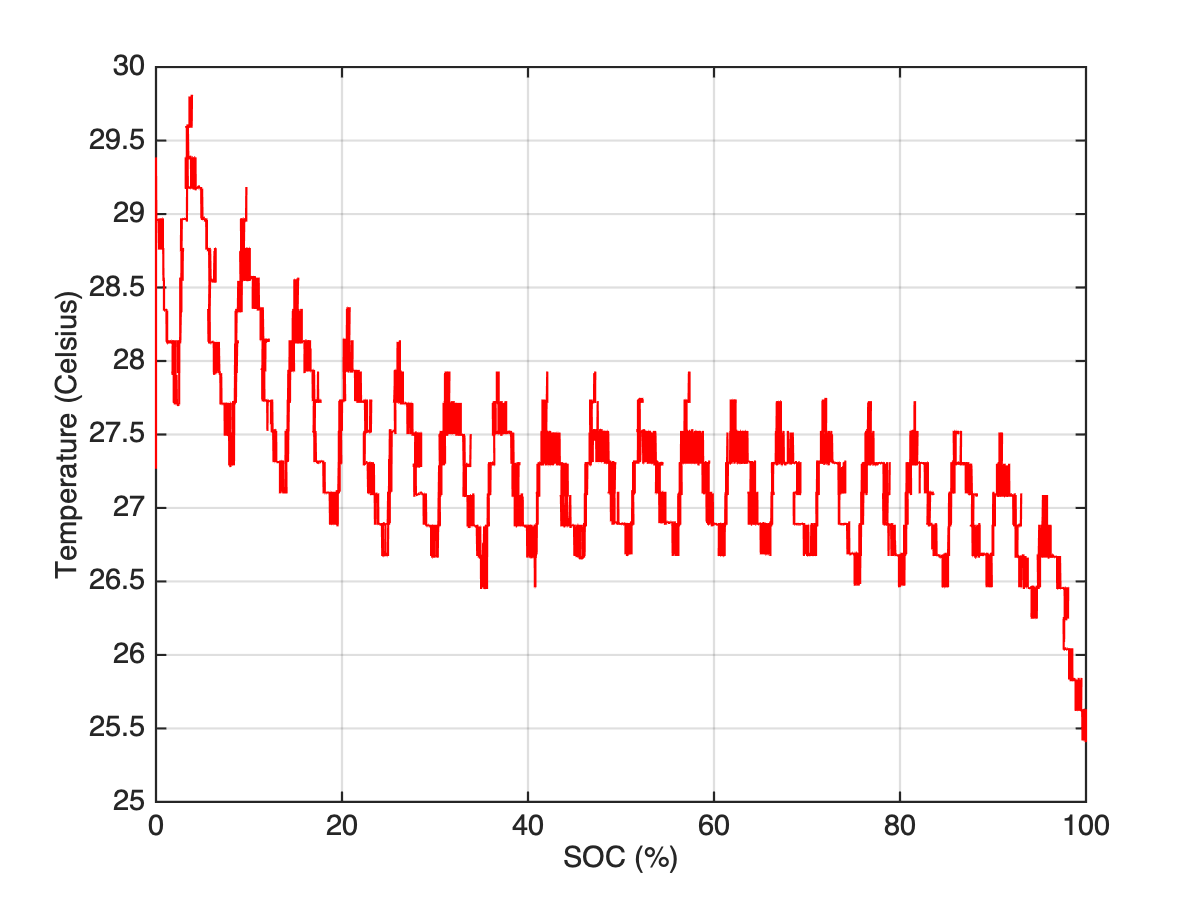}}		
	%\hspace{-1pt}%
	\subfloat[]{%
		\label{fig:Amphours-time}%
		\includegraphics*[scale=0.35]{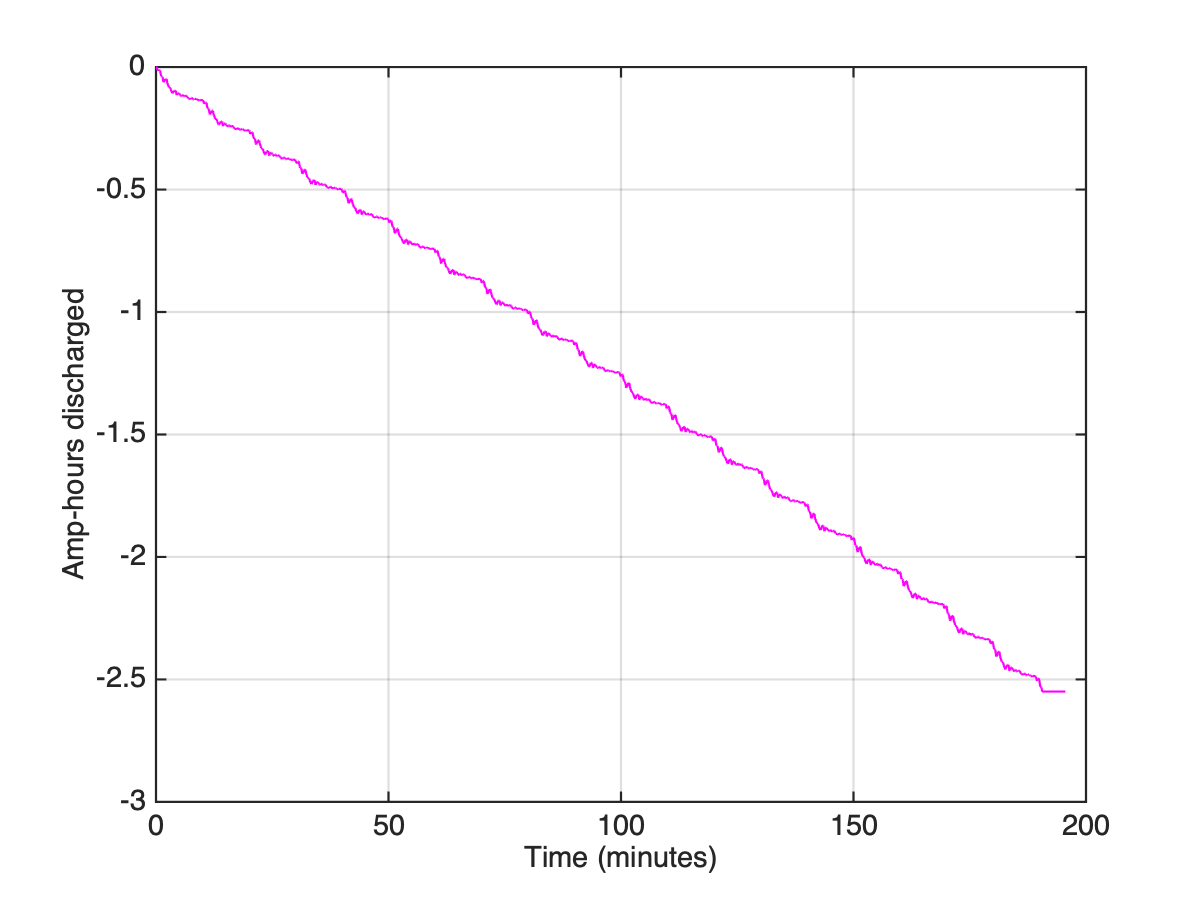}}\\
	\subfloat[]{%
		\label{fig:Voltage-time}%
		\includegraphics*[scale=0.35]{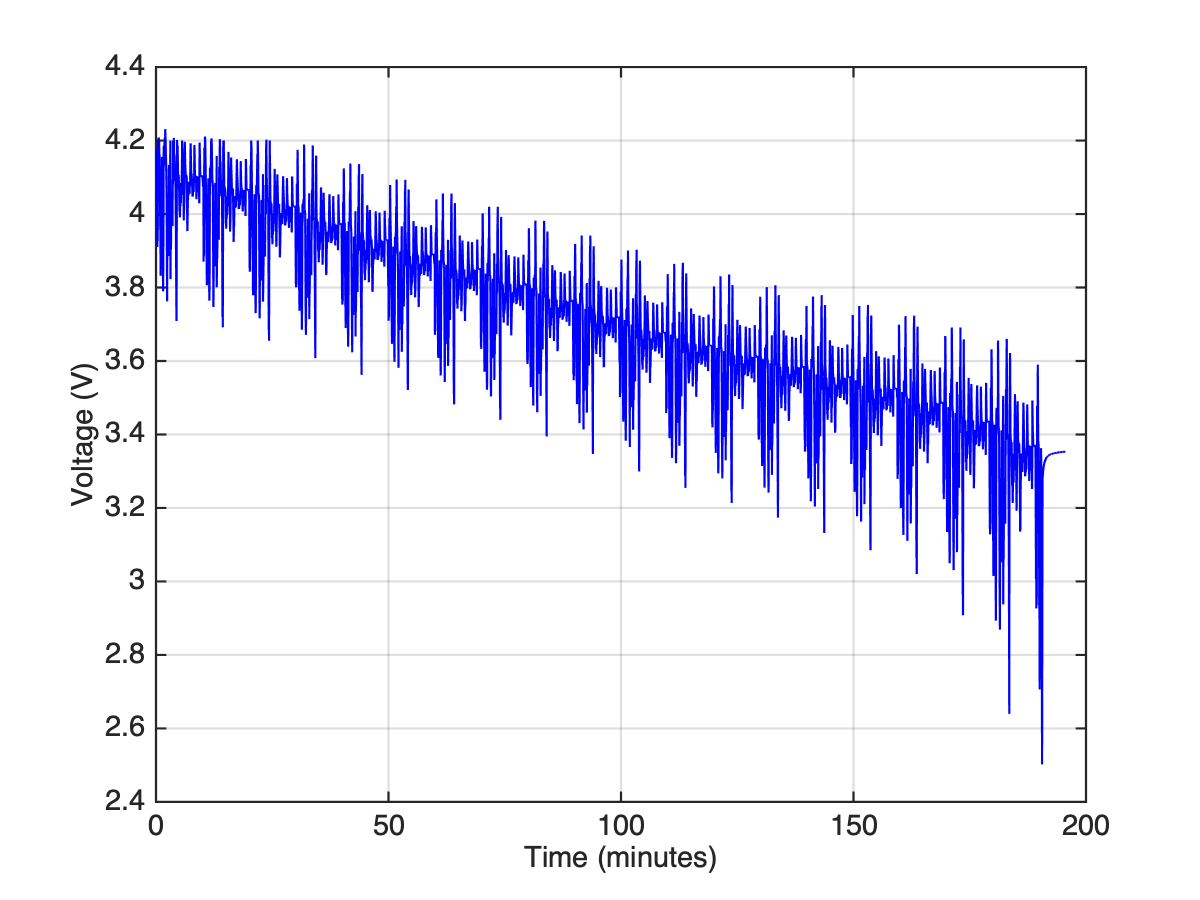}}
	\subfloat[]{%
		\label{fig:current-time}%
		\includegraphics*[scale=0.35]{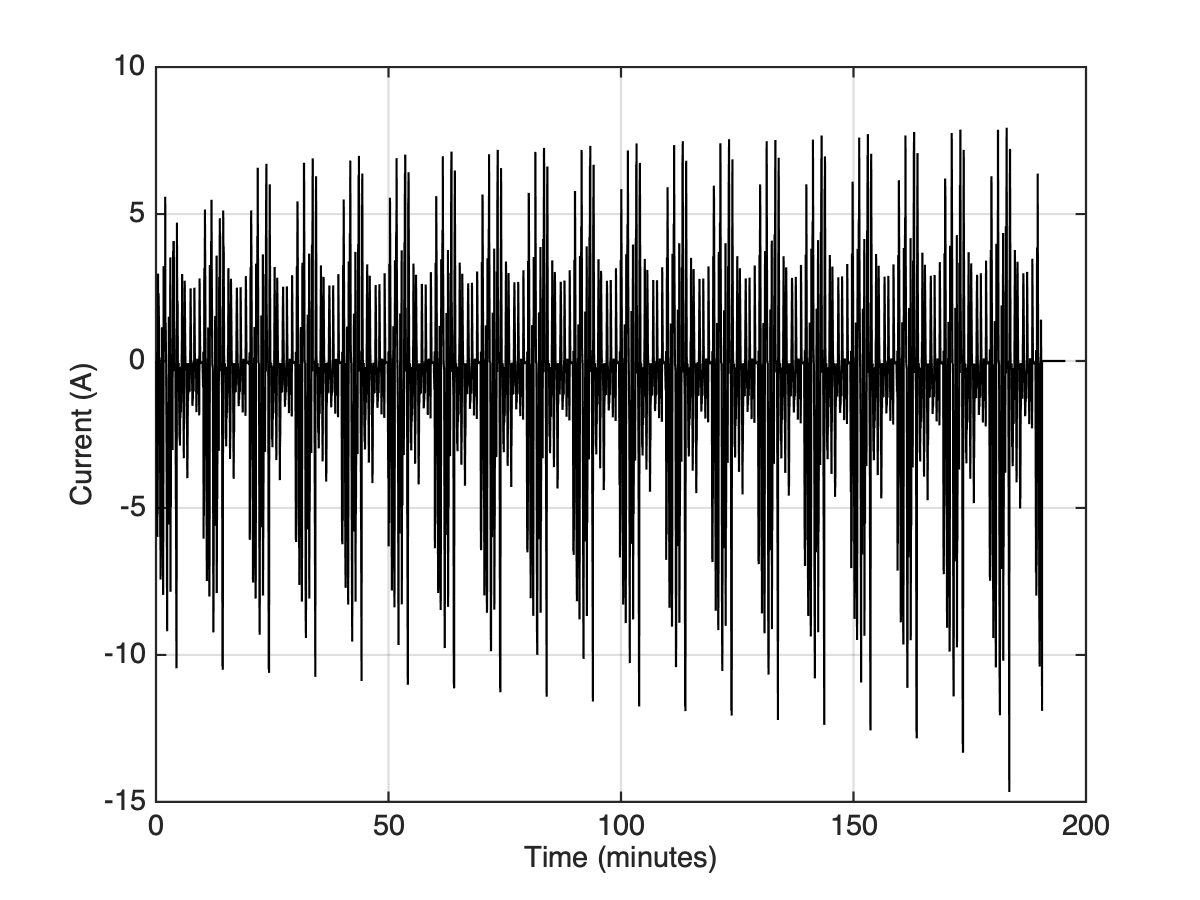}}\\
	\subfloat[]{%
		\label{fig:temperature-time}%
		\includegraphics*[scale=0.35]{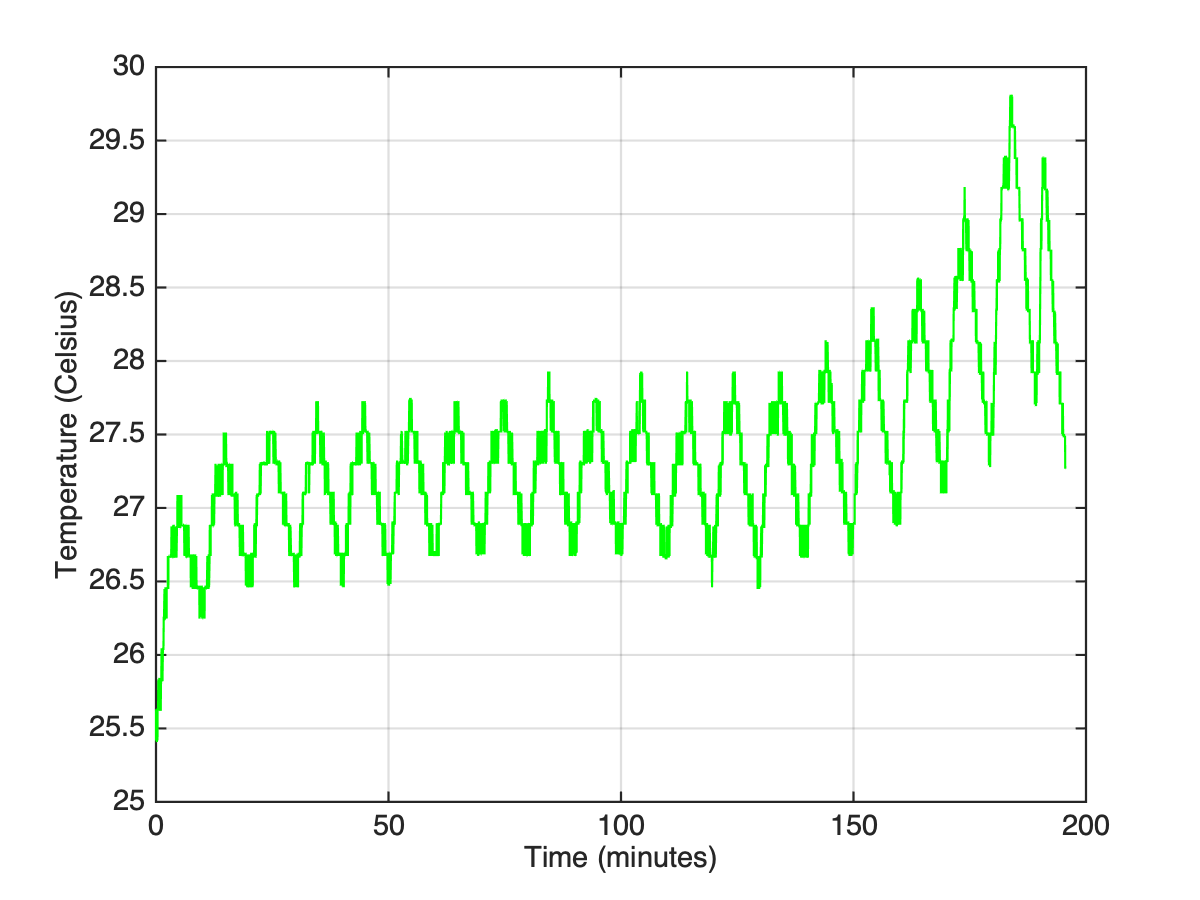}}
	\subfloat[]{%
		\label{fig:voltage-SOC}%
		\includegraphics*[scale=0.35]{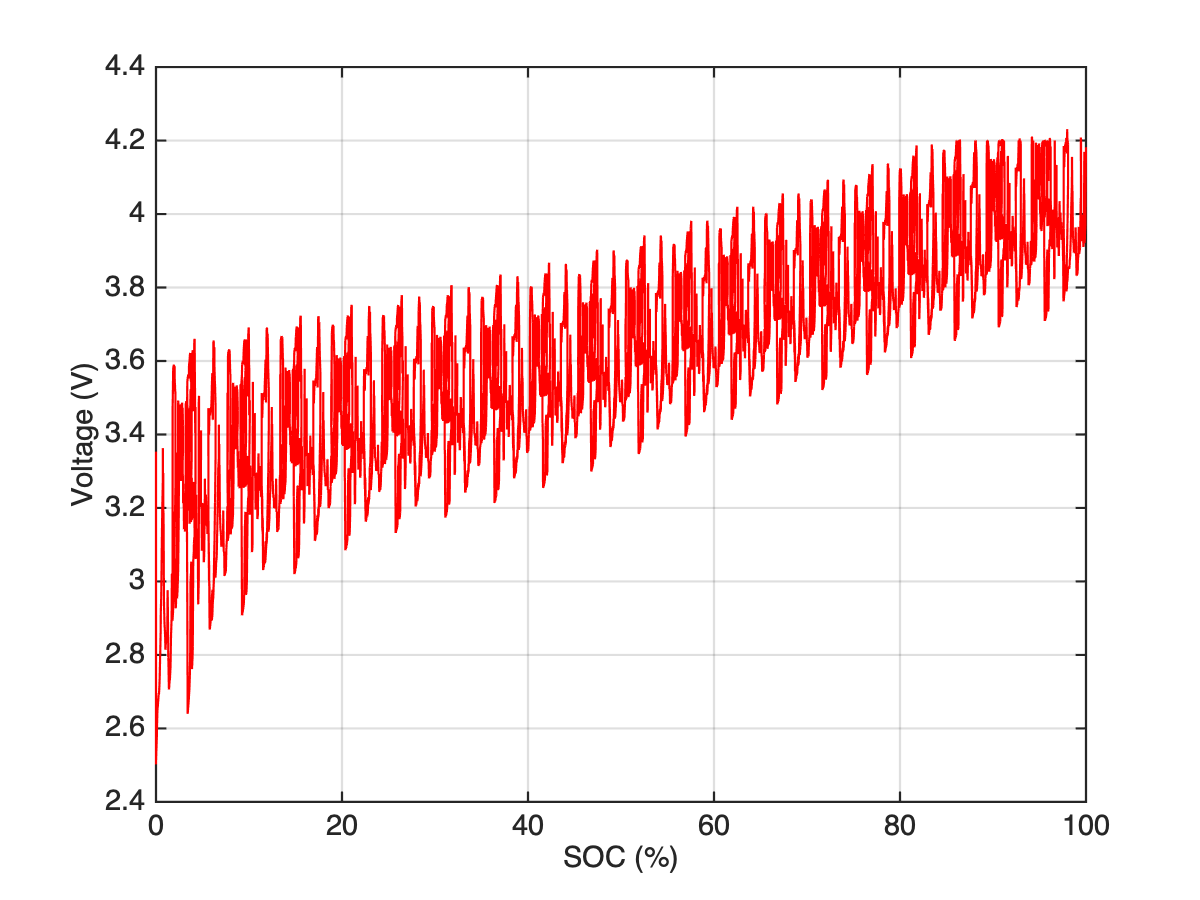}}\\
	\caption[]{
		in \subref{fig:temperature-SOC}, \subref{fig:Amphours-time}, \subref{fig:Voltage-time}, \subref{fig:current-time},
		\subref{fig:temperature-time}, and  \subref{fig:voltage-SOC} we have: temperature ($^o$ C) vs SOC (\%), amp-hours discharged vs time (minutes), voltage (V) vs time (minutes), current (A) vs time (minutes), temperature ($^o$ C) vs time (minutes), and voltage (V) vs SOC (\%), respectively.}%
	\label{fig:curvas-caracteristicas}%
\end{figure}

Fig. \ref{fig:curvas-caracteristicas} shows the following 2.9 Ah Panasonic 18650PF Li-ion cell characteristic curves:

\begin{itemize}
	\item temperature ($^o$ C) vs SOC (\%);
	\item amp-hours discharged vs time (minutes);
	\item voltage (V) vs time (minutes);
	\item current (A) vs time (minutes);
	\item temperature ($^o$ C) vs time (minutes); and
	\item voltage (V) vs SOC (\%).
\end{itemize}

The input data ($\rvx$) or features are: $x_1=v(t)$ (voltage in V), $x_2=i(t)$ (current in A), and $x_3=T(t)$ (temperature  in $^o$C), where $t$ denotes time in seconds. We see no reason for the inclusion of extra features in the hypothesis space, as the other data collected in \cite{Kollmeyer2018} are: i ) Wh (measured watt-hours, with Wh counter reset after each charge, test, or drive cycle), 
Power (measure power in watts), and 
Chamber-Temp-degC (measured chamber temperature in degrees Celsius). 
%as we are following the principle of parsimony, now most widely known as the Occam's razor. 
The output variable or target ($\rvy$) is the SOC (\%). The dataset has $116,982$ examples, which were divided examples for training, validation, and testing, respectively.

We applied feature normalization on the input data using the formula

\begin{equation}
\label{eq:feat-norm}
	\rvx_{\text{normalized}}  = \frac{\rvx - \mu_{\rvx}}{\sigma_{\rvx}} 
\end{equation} 

where $\mu_{\rvx}$ and $\sigma_{\rvx}$ denote the mean and standard deviation of $\rvx $.

We choose Mean Absolute Error (MAE) as the performance metric \cite{chemali2018} and the K-fold cross-validation as the basic method to fight overfitting. Notwithstanding, we have also used weight regularizers and dropout layers with the same goal.

The literature review indicated that the \texttt{Python} language is the most suitable for this research, for it has several free and open source frameworks for deep learning. In addition, \texttt{Python} is the language most used by the machine learning community \cite{cholletPy, gulli2019, skansi2018}.  Other options like \texttt{R} also  have great machine learning libraries \cite{allaire2017}.

Open source deep learning frameworks such as TensorFlow \cite{tf}, PyTorch \cite{pytorch}, MXNet \cite{mxnet} and Microsoft Cognitive Toolkit (CNTK) \cite{cntk} \footnote {The Caffe2 \cite{caffe2} library has been absorbed by PyTorch.} have stood out in recent years. However, TensorFlow 2 and the Keras API offer some differentials, such as the possibility of executing codes in Google Collaboratory, or "Colab" \cite{colab}, just using a browser and with free access to CPU/GPU (and even to Google's Tensor Processor Unit (TPU)).

In addition, TensorFlow offers a browser-based visualization tool called TensorBoard, whose main objective is to help the user visually monitor everything that happens inside their model during training \cite{cholletPy}. TensorBoard automates some features such as visualizing the learning curve of neural networks.

Thus, we decided to use the \texttt{Python} language and the TensorFlow 2 framework in conjunction with the Keras API. We coded/prototyped the deep learning model in the Spyder IDE (\texttt{Python} 3.7) of the Anaconda package manager. % and used Colab to speed up prototyping using GPU/TPU, which significantly reduced code execution time.

%Além disso, o TensorFlow oferece uma ferramenta de visualização baseada em browser denominada TensorBoard, cujo principal objetivo é ajudar o usuário a monitorar visualmente tudo o que acontece dentro do seu modelo durante o treinamento \cite{cholletPy}. O TensorBoard automatiza algumas funcionalidades como a visualização da curva de aprendizado das redes neurais.

%Assim, optou-se pela utilização da linguagem \texttt{Python} e pelo framework TensorFlow 2 em conjunto com a API Keras. 

%According to \cite{cholletPy}, 

%\begin{quotation}
%``Keras has the following key features
%	\begin{itemize}
%		\item It allows the same code to run seamlessly on CPU or GPU.
%		\item It has a user-friendly API that makes it easy to quickly prototype deep-learning
%		models.
%		\item It has built-in support for convolutional networks (for computer vision), recurrent networks (for sequence processing), and any combination of both.
%		\item It supports arbitrary network architectures: multi-input or multi-output models, layer sharing, model sharing, and so on. This means Keras is appropriate for building essentially any deep-learning model, from a generative adversarial net-
%		work to a neural Turing machine \cite{goodfellow2014generative, graves2014}.'' 
%	\end{itemize}
%\end{quotation} 

\subsection{Tuning of Hyperparameters}\label{subsec:tuning}

\begin{figure}[h]
	\centering
	\subfloat[]{
		\label{fig:DFN4}
		\includegraphics*[scale=0.2]{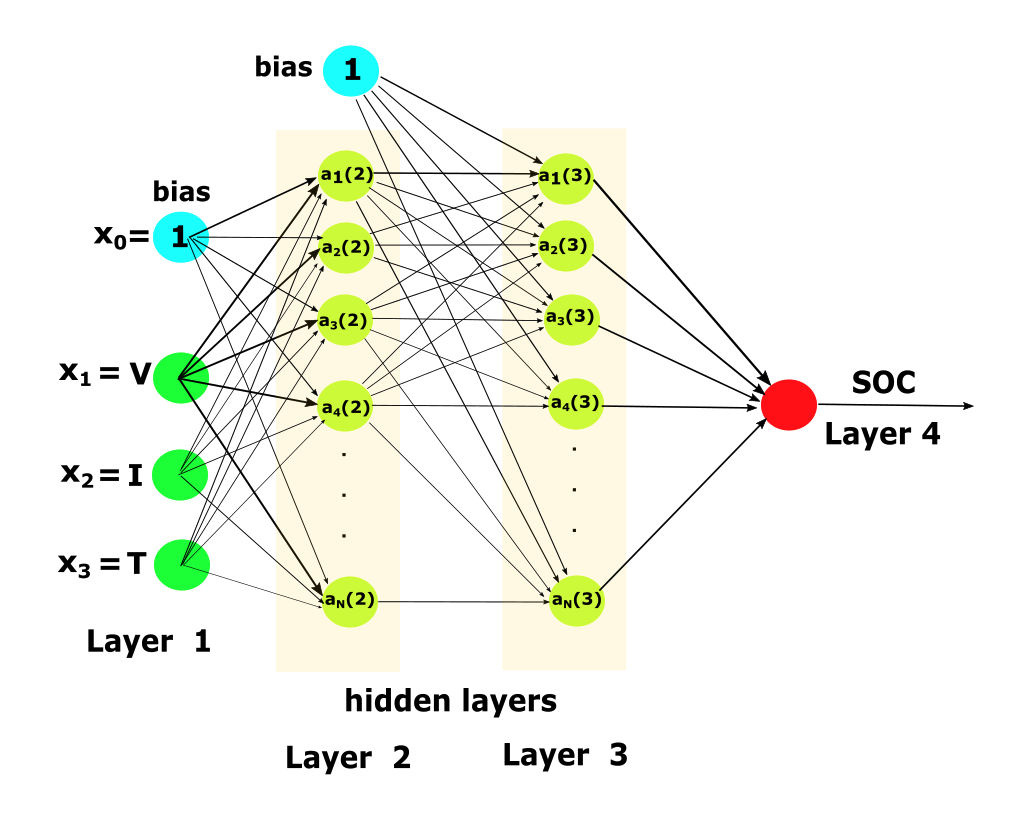}}
	\subfloat[]{
		\label{fig:DFN4-scheme}
		\includegraphics*[scale=0.2]{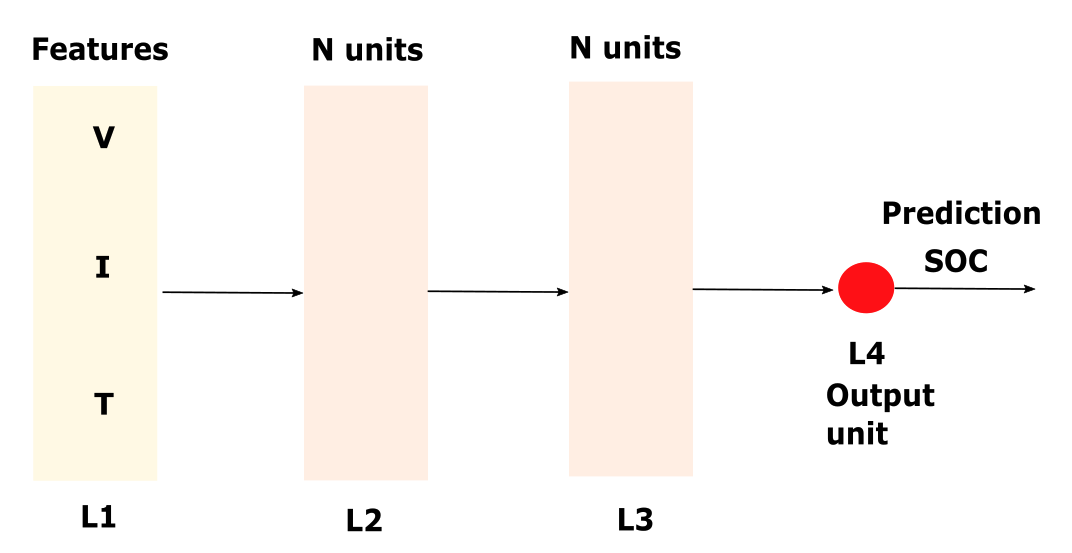}}
	\caption[]{
		\subref{fig:DFN4} and \subref{fig:DFN4-scheme}: densely connected 4-layer DFN and its schematic version,  respectively. }%
	\label{fig1:DFN}
\end{figure}

\begin{figure}[h]
	\centering
	\subfloat[]{
		\label{fig1:64batch-100epochs}
		\includegraphics*[height=4.3cm,width=6.2cm]{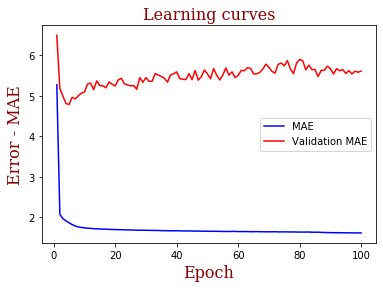}}		
	\subfloat[]{
		\label{fig2:128batch-50epochs}
		\includegraphics*[height=4.3cm,width=6.2cm]{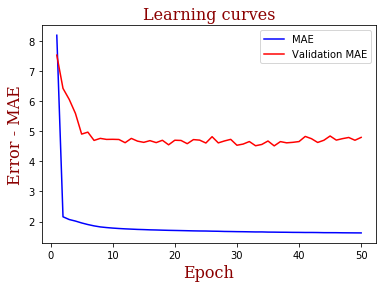}}\\
	\subfloat[]{%
		\label{fig3:256batch-30epochs}%
		\includegraphics*[height=4.3cm,width=6.2cm]{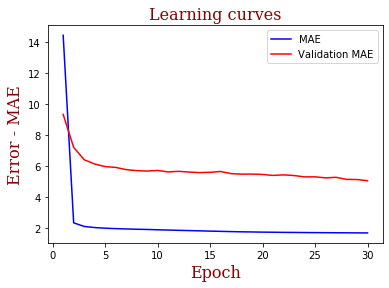}}\\
	\caption[]{
		in \subref{fig1:64batch-100epochs}, \subref{fig2:128batch-50epochs}, and \subref{fig3:256batch-30epochs}, we have learning curves for: $256$ units/hidden layer, batch size of $64$ and $100$ epochs; $256$ units/hidden layer,  batch size of $128$ and $50$ epochs; and $256$ units/hidden layer,  batch size of $256$, and $30$ epochs,  respectively. }%
	\label{fig:MAE-overfitting}%
\end{figure}

Fig. \ref{fig1:DFN} shows the architecture of a densely connected DFN with four (4) layers (two hidden layers).  Fig. \ref{fig:MAE-overfitting} shows the learning curves using Adam optimizer and $4$-fold cross-validations. Note that overfitting manifests itself as a gap between the validation MAE (red plot) and the training MAE (blue plot) in those figures. 

We tuned the hyperparameters of the deep learning model using  $L^1$/$L^2$ parameter norm penalties and adding two extra dropout layers.

In $L^2$ regularization, the cost added to the objective function is proportional to the square root of the sum of the square values of the weight coefficients ($L^2$  norm -- $||\rvw||^2_2$), whereas  in $L^1$ regularization the cost added to the objetive function is proportional to the sum of the absolute values of the weight coefficients ($L^1$ norm -- $||\rvw||_1$). 

\begin{figure}[h]
	\begin{center}
		\includegraphics*[scale = 0.2]{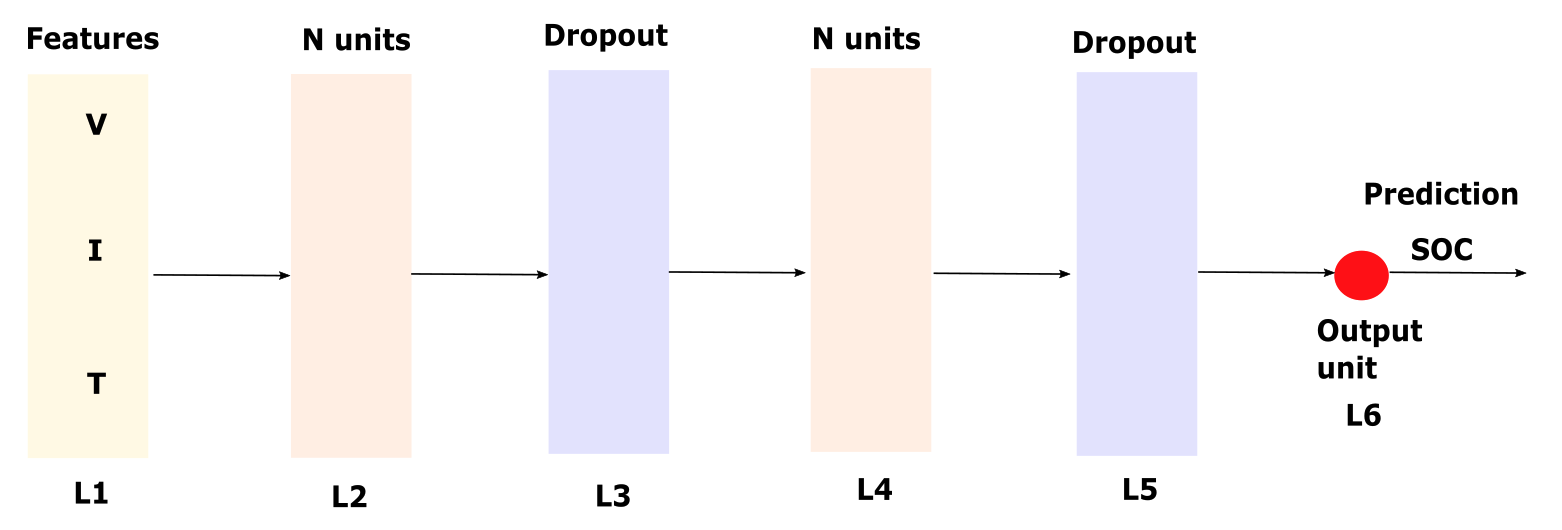}
	\end{center}	
	\caption{DFN with six (6) layers.}
	\label{fig2:DFN6}
\end{figure}

\begin{figure}[h]
	\begin{center}\includegraphics*[scale = 0.5]{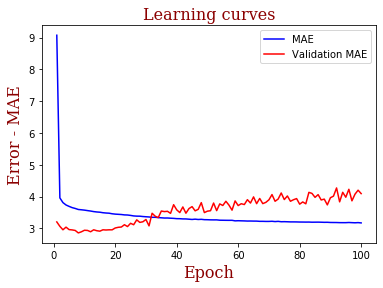}
	\end{center}	
	\caption{learning curves for a model with $256$ units/hidden layer,  batch size of $128$, dropout rate of $0.5$.}
	\label{fig1:MAE-overfitting}
\end{figure}

Fig. \ref{fig2:DFN6} shows the architecture of a DFN with six (6) layers, where we have, after the input layer (layer 1),  in sequence, two pairs of a $256$ units/hidden layer followed by a dropout  layer, then the final layer. The neural net uses $8$-fold cross-validations. In Fig. \ref{fig1:MAE-overfitting}, note that overfitting occurs only around $35$ epochs. Thus, this model has a greater power of generalization than the previous ones, as expected.

\subsection{SOC Estimation Results}\label{subsec:results}

The DFN model with two hidden layers, $256$ units/hidden layer, batch size of $128$ and without regularization achieves the 
best SOC's estimate on the test set, with  a MAE of approximately $1.60\%$.

However, as indicated by the learning curves in Fig. \ref{fig1:MAE-overfitting}, the DFN model with four hidden layers (two pairs of a $256$ units/hidden layer followed by a dropout layer with a dropout rate of $0.5$) has a greater power of generalization than the previous model. The MAE obtained on the test set was approximately $2.0\% $ in this case.
 
\section{Conclusions}\label{sec:conclusions}

This paper presents two simple DFN models with two and four hidden layers, respectively, using an optimizer with adaptive learning rules, and the Relu activation function, in order to estimate the State of Charge (SOC) of a Panasonic 18650PF lithium-ion battery of the Neural Network (NN) drive cycle of dataset \cite{Kollmeyer2018} using the K-fold cross-validation method. 

The DFN model with four hidden layers presents a better power of generalization, not only because it has a greater capacity in terms of more layers, but also due to the application of additional regularization techniques such as dropout layers and parameter norm penalties. 
%Our deep learning model achieves a Mean Absolute Error (MAE) of $1.6\%$ on the test set. 
The contribution of this work is to present a methodology of building a DFN for a lithium-ion battery and its performance assessment, which follows the best practices in machine learning.

%In future work we intend to:
%\begin{itemize}
%	\item investigate the modeling with recurrent neural networks (RNN);
%	\item estimate the SOC of the NN drive cycle of dataset \cite{Kollmeyer2018} at different temperatures, with a deep learning model for each temperature; and
%	\item estimate the SOC of the NN drive cycle of dataset \cite{Kollmeyer2018} at different temperatures with only one deep learning model.
%\end{itemize}

\bibliographystyle{unsrt}
\bibliography{references}

\begin{thebibliography}{10}
\providecommand{\url}[1]{#1}
\csname url@rmstyle\endcsname
\providecommand{\newblock}{\relax}
\providecommand{\bibinfo}[2]{#2}
\providecommand\BIBentrySTDinterwordspacing{\spaceskip=0pt\relax}
\providecommand\BIBentryALTinterwordstretchfactor{4}
\providecommand\BIBentryALTinterwordspacing{\spaceskip=\fontdimen2\font plus
\BIBentryALTinterwordstretchfactor\fontdimen3\font minus
  \fontdimen4\font\relax}
\providecommand\BIBforeignlanguage[2]{{%
\expandafter\ifx\csname l@#1\endcsname\relax
\typeout{** WARNING: IEEEtran.bst: No hyphenation pattern has been}%
\typeout{** loaded for the language `#1'. Using the pattern for}%
\typeout{** the default language instead.}%
\else
\language=\csname l@#1\endcsname
\fi
#2}}

\bibitem{sandia}
S.~N. Laboratories, ``{DOE}/{EPRI} 2013 electricity storage handbook in
  collaboration with {NRECA},'' Sandia National Laboratories, USA, Tech. Rep.,
  2013.

\bibitem{hannan2016}
M.~A. Hannan and et~al., ``Review of {E}nergy {S}torage {S}ystems for
  {E}lectric {S}torage {V}ehicle {A}pplications: {I}ssues and {C}hallenges,''
  \emph{Renewable and Sustainable Energy Reviews}, vol. Vol. 69, pp. 771 --
  789, 2016.

\bibitem{whit2012}
M.~S. Whittingham, ``History, {E}volution, and {F}uture {S}tatus of {E}nergy
  {S}torage,'' \emph{Proceedings of the IEEE}, vol. Vol. 100, pp. 1518 -- 1534,
  2012.

\bibitem{luo2015}
X.~Luo and et~al, ``Overview of {C}urrent {D}evelopment in {E}lectrical
  {E}nergy {S}torage {T}echnologies and the {A}pplication {P}otential in
  {P}ower {S}ystem {O}peration,'' \emph{Applied Energy}, vol. Vol. 137, pp. 511
  -- 536, 2015.

\bibitem{byrne2017}
R.~H. Byrne and et~al, ``Energy {M}anagement and {O}ptimization {M}ethods for
  {G}rid {E}nergy {S}torage {S}ystems,'' \emph{I{EEE} {ACCESS}}, 2017.

\bibitem{motapon2017}
S.~N. Motapon and et~al., ``A {G}eneric {E}lectrothermal {L}i-ion {B}attery
  {M}odel for {R}apid {E}valuation of {C}ell {T}emperature {T}emporal
  {E}volution,'' \emph{I{EEE} Transactions on Industrial Electronics}, vol.
  Vol. 64, no.~2, pp. 998 -- 1007, February 2017.

\bibitem{huria2012}
T.~Huria and et~al., ``High fidelity electrical model with thermal dependence
  for characterization and simulation of high power lithium battery cells,'' in
  \emph{IEEE International Electric Vehicle Conference}.\hskip 1em plus 0.5em
  minus 0.4em\relax IEEE, 2012.

\bibitem{ren2018}
L.~Ren and et~al, ``Remaining useful life prediction for lithium-ion battery: A
  deep learning approach,'' \emph{I{EEE} {ACCESS}}, vol.~6, pp.
  50\,587--50\,598, 2018.

\bibitem{jeon2014}
D.~H. Jeon, ``{N}umerical {M}odeling of {L}ithium {I}on {B}attery for
  {P}redicting {T}hermal {B}ehavior in a {C}ylindrical {C}ell,'' \emph{Current
  Appl. Phys.}, vol. Vol. 14, no.~2, pp. 196 -- 205, Feb. 2014.

\bibitem{li2014}
J.~Li and et~al., ``{A}n {E}lectrochemical-{T}hermal {M}odel {B}ased on
  {D}ynamics {R}esponses for {L}ithium {I}ron {P}hosphate {B}attery,'' \emph{J.
  Power Sources}, vol. Vol. 255, pp. 130 -- 143, June 2014.

\bibitem{zhao2017}
R.~{Zhao}, P.~J. {Kollmeyer}, R.~D. {Lorenz}, and T.~M. {Jahns}, ``A compact
  unified methodology via a recurrent neural network for accurate modeling of
  lithium-ion battery voltage and state-of-charge,'' in \emph{2017 IEEE Energy
  Conversion Congress and Exposition (ECCE)}, 2017, pp. 5234--5241.

\bibitem{chemali2018}
E.~Chemali and et~al, ``State-of-charge estimation of li-ion batteries using
  deep neural networks: a machine learning approach,'' \emph{Journal of {P}ower
  {S}ources}, vol. 400, pp. 242--255, 2018.

\bibitem{chen2019}
B.~Chen, T.~Medini, and A.~Shrivastava, ``S{LIDE}: {I}n {D}efense of {S}mart
  {A}lgorithms over {H}ardware {A}cceleration for {L}arge-{S}cale {D}eep
  {L}earning {S}ystems,'' \emph{arXiv}, march 2019.

\bibitem{goodf2016}
I.~GoodFellow, Y.~Bengio, and A.~Courville, \emph{Deep {L}earning},
  1st~ed.\hskip 1em plus 0.5em minus 0.4em\relax MIT Press, 2016.

\bibitem{lecun2015}
Y.~LeCun, Y.~Bengio, and G.~Hinton, ``Deep learning,'' \emph{Nature}, vol. 521,
  pp. 436--444, 2015.

\bibitem{mallat2016}
\BIBentryALTinterwordspacing
S.~Mallat, ``Understanding deep convolutional networks,'' \emph{Phil. Trans. R.
  Soc. A}, 2016. [Online]. Available:
  \url{http://rsta.royalsocietypublishing.org}
\BIBentrySTDinterwordspacing

\bibitem{shervin2019}
\BIBentryALTinterwordspacing
S.~Minaee and A.~A. Abdolrashidi, ``Deep-emotion: Facial expression recognition
  using attentional convolutional network,'' \emph{arXiv}, february 2019.
  [Online]. Available: \url{http://arxiv.org/abs/1902.01019}
\BIBentrySTDinterwordspacing

\bibitem{silver2018}
D.~Silver and et. al., ``Mastering chess and shogi by sef-play with a general
  reinforcement learning algorithm,'' \emph{Science}, vol. 362, pp. 1140--1144,
  dec. 2018.

\bibitem{lstm-li-ion-2018}
E.~{Chemali}, P.~J. {Kollmeyer}, M.~{Preindl}, R.~{Ahmed}, and A.~{Emadi},
  ``Long short-term memory networks for accurate state-of-charge estimation of
  li-ion batteries,'' \emph{IEEE Transactions on Industrial Electronics},
  vol.~65, no.~8, pp. 6730--6739, 2018.

\bibitem{Kollmeyer2017}
P.~{Kollmeyer}, A.~{Hackl}, and A.~{Emadi}, ``Li-ion battery model performance
  for automotive drive cycles with current pulse and eis parameterization,'' in
  \emph{2017 IEEE Transportation Electrification Conference and Expo (ITEC)},
  2017, pp. 486--492.

\bibitem{sepasi2014}
S.~Sepasi, R.~Ghorbani, and B.~Y. Liaw, ``Improved extended kalman filter for
  state of charge estimation of battery pack,'' \emph{Journal of Power
  Sources}, vol. 255, pp. 368--376, 2014.

\bibitem{chark2010}
M.~Charkhgard and M.~Farrokhi, ``State-of-charge estimation for lithium-ion
  batteries using neural networks and ekf,'' \emph{{IEEE} Transactions on
  Industrial Electronics}, vol.~57, dec. 2010.

\bibitem{du2014}
J.~Du, Z.~Liu, and Y.~Wang, ``State of charge estimation for li-ion battery
  based on model from extreme learning machine.'' \emph{Control Engineering
  Practice}, vol.~26, pp. 11--19, 2014.

\bibitem{lee2014}
Y.-S. Lee, W.-Y. Wang, and T.-Y. Kuo, ``Soft computing for battery
  state-of-charge (bsoc) estimation in battery string systems,'' \emph{I{EEE}
  Transactions on Industrial Electronics}, vol.~55, no.~1, jan. 2014.

\bibitem{chang2013}
W.-Y. Chang, ``Estimation of the state of charge for a lfp battery using a
  hybrid method that combines a rff neural network, an ols algorithm and aga,''
  \emph{Electrical Power and Energy Systems}, vol.~53, pp. 603-- 611, 2013.

\bibitem{cholletPy}
F.~Chollet, \emph{Deep {L}earning with {P}ython}, 1st~ed.\hskip 1em plus 0.5em
  minus 0.4em\relax Manning Publications, 2018.

\bibitem{kevin2012}
K.~P. Murphy, \emph{Machine Learning: A Probabilistic Approach}.\hskip 1em plus
  0.5em minus 0.4em\relax The MIT Press, 2012.

\bibitem{handbook-AI-vol1}
{R. Anderson et al}, ``{Chapter I -- Introduction},'' in \emph{{The Handbook of
  Artificial Intelligence}}, A.~Barr and E.~A. Feigenbaum, Eds., vol.~1.\hskip
  1em plus 0.5em minus 0.4em\relax Elsevier, 1981.

\bibitem{boden2016}
M.~A. Boden, \emph{{AI Its Nature and Future}}, 1st~ed.\hskip 1em plus 0.5em
  minus 0.4em\relax Oxford University Press, 2016.

\bibitem{pitts1943}
\BIBentryALTinterwordspacing
W.~S. McCulloch and W.~Pitts, ``A logical calculus of the ideas immanent in
  nervous activity,'' \emph{Bulletin of Mathematical Biophysics}, vol.~5, pp.
  115--133, 1943. [Online]. Available:
  \url{http://www.cs.cmu.edu/~epxing/Class/10715/reading/McCulloch.and.Pitts.pdf}
\BIBentrySTDinterwordspacing

\bibitem{turing1936}
A.~M. Turing, ``On {C}omputable {N}umbers, with an {A}pplication to the
  {E}ntscheidungsproblem,'' \emph{Proceedings of the London Mathematical
  Society}, vol.~42, pp. 230--265, 1936.

\bibitem{russel}
\BIBentryALTinterwordspacing
S.~Russel and P.~Norvig, \emph{Intelig\^encia Artificial}, terceira
  ed.~ed.\hskip 1em plus 0.5em minus 0.4em\relax Rio de Janeiro: Elsevier,
  2013. [Online]. Available: \url{http://aima.cs.berkeley.edu/}
\BIBentrySTDinterwordspacing

\bibitem{cybenko1989}
\BIBentryALTinterwordspacing
G.~Cybenko, ``Approximation by superpositions of a sigmoidal function,''
  \emph{Math. Control Signal Systems}, vol.~2, pp. 303--314, 1989. [Online].
  Available: \url{https://doi.org/10.1007/BF02551274}
\BIBentrySTDinterwordspacing

\bibitem{opp2009}
A.~V. Oppenheim and R.~W. Schafer, \emph{{Discrete-Time Signal Processing}},
  3rd~ed.\hskip 1em plus 0.5em minus 0.4em\relax Pearson, 2009.

\bibitem{lathi1998}
B.~P. Lathi, \emph{{Signal Processng and Linear Systems}}.\hskip 1em plus 0.5em
  minus 0.4em\relax Oxford University Press, 1998.

\bibitem{newell1980}
A.~Newell, ``Physical symbom systems,'' \emph{Cognitive Science}, vol.~4, pp.
  135--183, April--June 1980.

\bibitem{acm2018}
{ACM}, ``{A. M. Turing Award 2018},'' https://awards.acm.org/about/2018-turing,
  2020.

\bibitem{Cun90handwrittendigit}
L.~Cun, B.~Boser, J.~S. Denker, D.~Henderson, R.~E. Howard, W.~Hubbard, and
  L.~D. Jackel, ``{Handwritten Digit Recognition with a Back-Propagation
  Network},'' in \emph{Advances in Neural Information Processing
  Systems}.\hskip 1em plus 0.5em minus 0.4em\relax Morgan Kaufmann, 1990, pp.
  396--404.

\bibitem{McCarthy1990}
\BIBentryALTinterwordspacing
J.~McCarthy and E.~A. Feigenbaum, ``{In Memoriam: Arthur Samuel: Pioneer in
  Machine Learning},'' \emph{AI Magazine}, vol.~11, no.~3, p.~10, Sep. 1990.
  [Online]. Available:
  \url{https://www.aaai.org/ojs/index.php/aimagazine/article/view/840}
\BIBentrySTDinterwordspacing

\bibitem{samuel1959}
A.~L. {Samuel}, ``{Some Studies in Machine Learning Using the Game of
  Checkers},'' \emph{IBM Journal of Research and Development}, vol.~3, no.~3,
  pp. 210--229, 1959.

\bibitem{ertel2017}
W.~Ertel, \emph{Introduction to Artificial Intelligence}, 2nd~ed.\hskip 1em
  plus 0.5em minus 0.4em\relax Springer, 2017.

\bibitem{tom1997}
T.~M. Mitchell, \emph{Machine Learning}.\hskip 1em plus 0.5em minus 0.4em\relax
  McGraw-Hill Dscience/Eng./Math, 1997.

\bibitem{alphazero}
\BIBentryALTinterwordspacing
D.~Silver and et~al, ``General {R}einforcement {L}earning {A}lgorithm,''
  \emph{arXiv}, 2017. [Online]. Available:
  \url{https://arxiv.org/pdf/1712.01815.pdf?utm_campaign=nathan.ai%20newsletter&utm_medium=email&utm_source=Revue%20newsletter}
\BIBentrySTDinterwordspacing

\bibitem{widrow1960}
B.~Widrow and M.~E. Hoff, ``Adaptive switching circuits,'' \emph{IRE WESCON
  Convention Record}, pp. 96--104, 1960.

\bibitem{rosenblatt1958}
F.~Rosenblatt, ``The {P}erceptron: a probabilistic model for information
  storage and organization in the brain,'' \emph{Psychological Review},
  vol.~65, pp. 386--408, 1958.

\bibitem{hinton1986}
D.~E. Rumelhart, G.~E. Hinton, and J.~W. Ronald, ``{Learning Internal
  Representations by Error Propagation},'' in \emph{{Parallel Distributed
  Processing: Explorations in The Microstructure of Cognition}}, D.~E.
  Rumelhart and J.~L. McClelland, Eds., vol.~1.\hskip 1em plus 0.5em minus
  0.4em\relax Foundations, Cambridge, MA: Bradford Books/MIT Press, 1986.

\bibitem{bryson}
A.~E.~B. Jr. and Y.~C. Ho, \emph{{Applied Optimal Control}}.\hskip 1em plus
  0.5em minus 0.4em\relax Blaisdell, 1969.

\bibitem{werbos74}
P.~J. Werbos, ``{Beyond Regression: New tools for prediction and analysis in
  the behavioral sciences},'' 1974.

\bibitem{parker85}
D.~B. Parker, ``{Learning Logic},'' Center for Computational Research in
  Economics and Management Science, MIT, 1985.

\bibitem{lecun1986}
Y.~LeCun, ``{A Learning Scheme for Asymmetric Threshold Network},'' in
  \emph{{Disordered systems and biological organization}}, E.~Bienenstock,
  F.~Fogelman, and G.~Weisbuch, Eds.\hskip 1em plus 0.5em minus 0.4em\relax
  Springer Verlag, 1986.

\bibitem{papoulis91}
A.~Papoulis, \emph{Probability, Random Variables, and Stochastic Processes},
  3rd~ed.\hskip 1em plus 0.5em minus 0.4em\relax McGraw-Hill, 1996.

\bibitem{wolpert1996}
D.~H. Wolpert, ``{The Lack of A Priori Disctintions Between Learning
  Algorithms},'' \emph{Neural Computation}, vol.~8, pp. 1341--1390, 1996.

\bibitem{hinton2006}
G.~E. Hinton, S.~Osindero, and Y.~Teh, ``A fast learning algorithm for deep
  belief nets,'' \emph{Neural Computation}, vol.~18, pp. 1527--1554, 2006.

\bibitem{allaire2017}
F.~Chollet and J.~J. Allaire, \emph{{Deep Learning with R}}, 1st~ed.\hskip 1em
  plus 0.5em minus 0.4em\relax Manning Publications, 2017.

\bibitem{Kollmeyer2018}
\BIBentryALTinterwordspacing
P.~Kollmeyer, ``{Panasonic 18650PF Li-ion Battery Data},'' \emph{Mendeley
  Data}, vol.~1, 2018. [Online]. Available:
  \url{https://data.mendeley.com/datasets/wykht8y7tg/1}
\BIBentrySTDinterwordspacing

\bibitem{gulli2019}
A.~Gulli, A.~Kapoor, and S.~Pal, \emph{Deep {L}earning with {T}ensor{F}low 2
  and {K}eras}, 2nd~ed.\hskip 1em plus 0.5em minus 0.4em\relax Packt>, 2019.

\bibitem{skansi2018}
S.~Skansi, \emph{{Introduction to Deep Learning: From Logical Calculus to
  Artificial Intelligence}}, 1st~ed.\hskip 1em plus 0.5em minus 0.4em\relax
  Springer, 2018.

\bibitem{tf}
\BIBentryALTinterwordspacing
Google, ``Tensorflow,'' 2020. [Online]. Available:
  \url{https://www.tensorflow.org/?hl=pt-br}
\BIBentrySTDinterwordspacing

\bibitem{pytorch}
\BIBentryALTinterwordspacing
A.~Paszke, S.~Gross, S.~Chintala, and G.~Chanan, ``Pytorch,'' 2020. [Online].
  Available: \url{https://pytorch.org}
\BIBentrySTDinterwordspacing

\bibitem{mxnet}
\BIBentryALTinterwordspacing
Apache, ``{MXN}et,'' 2020. [Online]. Available: \url{https://mxnet.apache.org}
\BIBentrySTDinterwordspacing

\bibitem{cntk}
\BIBentryALTinterwordspacing
Microsoft, ``Microsoft {C}ognitive {T}oolkit,'' 2020. [Online]. Available:
  \url{https://docs.microsoft.com/en-us/cognitive-toolkit/}
\BIBentrySTDinterwordspacing

\bibitem{caffe2}
\BIBentryALTinterwordspacing
Facebook, ``Caffe2,'' 2020. [Online]. Available: \url{https://caffe2.ai}
\BIBentrySTDinterwordspacing

\bibitem{colab}
\BIBentryALTinterwordspacing
Google, ``Colaboratory,'' 2020. [Online]. Available:
  \url{https://colab.research.google.com/notebooks/intro.ipynb}
\BIBentrySTDinterwordspacing

\end{thebibliography}

\end{document}